# CENTER MANIFOLD AND MULTIVARIABLE APPROXIMANTS APPLIED TO NON-LINEAR STABILITY ANALYSIS

J-J. SINOU*, F. THOUVEREZ and L. JEZEQUEL.

Laboratoire de Tribologie et Dynamique des Systèmes UMR CNRS 5513,
Ecole Centrale de Lyon, 36 avenue Guy de Collongue, 69134 Ecully Cedex, France.

## ABSTRACT

This paper presents a research devoted to the study of instability phenomena in non-linear model with a constant brake friction coefficient. This paper outlines the stability analysis and a procedure to reduce and simplify the non-linear system, in order to obtain limit cycle amplitudes. The center manifold approach, the multivariable approximants theory, and the alternate frequency/time domain (AFT) method are applied. Brake vibrations, and more specifically heavy trucks grabbing are concerned. The modelling introduces sprag-slip mechanism based on dynamic coupling due to buttressing. The non-linearity is expressed as a polynomial with quadratic and cubic terms. This model does not require the use of brake negative coefficient, in order to predict the instability phenomena. Finally, the center manifold approach, the multivariable approximants, and the AFT method are used in order to obtain equations for the limit cycle amplitudes. These methods allow the reduction of the number of equations of the original system in order to obtain a simplified system, without loosing the dynamics of the original system, as well as the contributions of non-linear terms. The goal is the validation of this procedure for a complex non-linear model by comparing results obtained by solving the full system and by using these methods. The brake friction coefficient is used as an unfolding parameter of the fundamental Hopf bifurcation point.

## 1 INTRODUCTION

During the last years, the knowledge of the dynamic behaviour of systems with non-linear phenomena has been developed in order to exploit the full capability of structures by using systems in the non-linear range. Usually, the non-linear equations of motion are linearized at the steady-state operating point, and so a set of linearized perturbation equations is obtained. The stability was investigated by determining the eigenvalues of the linearized perturbation equations at each steady-state operating point. While stability analyses are extremely useful in evaluating the effect of changes in various system parameters, they cannot evaluate limit cycles amplitudes.
Of course, robust softwares have been developed in order to solve differential-algebraic equations corresponding to systems including several nonlinearities; and time history response solutions of the full set of non-linear equations can determine the vibration amplitude. But time history response solutions of the full set of non-linear equations are both time consuming and costly. For this reason, an understanding of the behaviour of systems having many degrees of freedom requires simplification methods in order to reduce the order of the system of equations, and/or eliminate as many nonlinearities as possible in the system of equations. Moreover, many physical systems are modeled by differential equations depending on a control parameter. In the study of the dynamic behaviour of such systems, bifurcation problems often arise within the control parameter range.
Due to the fact that such non-linear systems occur in many disciplines of engineering and science, considerable work has been devoted the development of methods for the approximation of the frequency response of non-linear systems, and so allowing explicit reductions. One of the most popular method for the approximation of the frequency response of non-linear systems is based on the balance of the harmonic components: the harmonic balance (HB) method (Nayfeh and Mook [1]), the incremental harmonic balance (IHB) method (Cheung, Chen and Lau [2], Leung and Chui [3], Lau and Zhang [4], Pierre, Ferri and Dowell [5]) and the alternate frequency/time domain (AFT) method (Cameron and Griffin [6] , Narayanan and Sekar [7]). Moreover, perturbation methods, such as the methods of multiple scales and averaging methods (Nayfeh and Balachandran [8]), have been used as simplification methods in many studies. There is a reduction in the dimension as one goes from the original system to the averaged system. Moreover, the normal form approach can also be used in order to eliminate as many non-linear terms of the non-linear equations as possible, through a non-linear change of the variables. These problems have already been studied by several researchers (Nayfeh and Mook [1], Brjuno[9]-[10], Guckhenheimer and Holmes [11], Jezequel and Lamarque[12], Iooss [13]-[14], Hsu[15]-[16] etc.). Moreover, one of the most important simplification method is the center manifold approach. The center manifold theorem (Marsden and McCracken [17]) characterises the local bifurcation analysis in the vicinity of a fixed point of the non-linear system. The center manifold approach reduces the original system to a center manifold associated with the part of the original system characterized by the eigenvalues with zero real parts at the bifurcation point. The center manifold may have smaller dimensions than that of the original system (Nayfeh [8], Guckhenheimer [11] and Knoblock [18]).
Another way to analyse and approximate the non-linear system is the theory of Padé approximants (Baker and Graves-Morris [19] and Brezinski [20]). The approximants are rational approximations for functions defined as a formal series expansion. Padé approximants have many applications in physics, but this technique is not yet usually used in the approximation of the non-linear problems.
In this paper, we propose to apply successively the center-manifold reduction, the multivariable approximants and the alternate frequency/time domain (AFT) method to the study of a self-excited system with many-degree-of-freedom, containing quadratic and cubic non-linear terms, characteristics of the modelling of heavy trucks grabbing. The center manifold reduction, the multivariable approximants, and the AFT method will be used in order to simplify and express the the original and final solution respectively of the last non-linear approximated system, and also to obtain an explicit description of the solution by their Fourier components.
We will first introduce some basic concepts of friction and brake noise. Next, we will present a model for analyses of grabbing mode vibration in automobile braking systems. The model does not use brake negative damping and predicts that system instability can occur with a constant brake friction coefficient. Finally, we will use the center manifold approach, multivariable approximants and the alternate frequency/time domain (AFT) method in order to predict limit cycle amplitudes. Results from center manifold approach, multivariable approximants, and AFT method will be compared with results obtained by the integration of the full original system, in order to validate this global procedure that employs successively, in a certain order, non-linear methods for the reduction and the simplification of the original system.

## 2 FRICTION INDUCED VIBRATION

A serious difficulty in the study of the stability analysis is due to the fact that the dynamic stability of a brake system depends on a number of factors such as friction coefficient, mechanical interaction, and stiffness for example. As a result, much effort has been done in the determination of models and mechanisms that predict friction induced vibrations. A lot of work on the brake noise and vibration was published during the last years. However there has been no uniformly accepted theory to characterize the problem, despite the investigation of various types of vibrations, such as disk brake squeal (Chambrette [21], North [22]), aircraft brake squeal (Liu and Ozbek [23]) and railway wheel squeal (Rudd [24]). In this way, analytical models have been proposed for the description of the dynamics of brake systems, including brake calliper, pads and disc: some of the more famous studies were proposed by Jarvis and Mills [25] (cantilever-disc models), Earles and Lee[26] (pin-disc models), and Spurr[27] (sprag-slip model).

One of the most important phases in studying the brake systems is the determination of the mechanism of the unstable friction induced vibration in brake systems. There is no unique mathematical model and theory for the explanation of the mechanisms and dynamic phenomena associated with friction. According to Ibrahim [28]-[29], Oden and Martins[30] and Crolla [31], there are four general mechanisms for friction-induced system instability, and more specifically friction-induced vibration in disc-brake systems: stick-slip, variable dynamic friction coefficient, sprag-slip, and coupling mechanism. The first two approaches rely on changes in the friction coefficient with relative sliding speed affecting the system stability. The last two approaches used kinematic constraints and modal coupling in order to develop the instability. Stick-slip is a low sliding speed phenomenon caused when the static friction coefficient is higher than the dynamic coefficient. A simple system that has been used to examine the stick-slip phenomenon is a mass sliding on a moving belt as shown in Figure 1(A)(a). Then, sliding and sticking occur in succession. The phenomenon associated with a friction coefficient decreases with rubbing speed is shown in Figure 1(A)(b). Due to this negative slope, the steady state sliding becomes unstable and caused friction-induced vibrations. Moreover, sprag-slip phenomenon occurs due to locking action of the slider into the sliding surface as defined in Figure 1(B). An important failure of this mechanism is the angle $\alpha$ between the resulting force at the friction contact and the normal direction of the sliding belt. Then, researchers gradually increased the sophistication of these sprag slip models by developing a more generalised theory, describing the mechanism as a geometrically induced or kinematic constraint instability.

In spite of numerous recent studies on the subject, the analysis of mechanism of disc brakes still presents a broad. Effectively, there are many types of brake vibration problems with various phenomena. Specialists as Crolla and Lang [31] divided them into three headings: disc brake noise, brake grabbing and brake drum noise.

Generally, brake noises are divided into categories according to the sound frequency. On the basis of previous brake experiments, there are many types of brake noises with various phenomena as squeal noise, groan noise, jerder noise, squelch noise, and pinch-out noise. Squeal noise and groan noise are the two important phenomena of brake noise. Technically speaking, noise is the result of a self-excited oscillation or dynamic instability of the brake. Squeal is accepted as being the result of such instabilities. For example, squeal can be due to a resonance of drums, rotors, or back plates; the frequency spectrum of squeal is in the 1 – 10 kHz range. In contrast to squeal, groan occurs at very slow vehicle speed. It is caused by stick-slip at the rubbing surface; the frequency spectrum of groan is in the 10 – 300 Hz range. The most important drum brake noise is squeal. As drum brakes were gradually replaced by disc brakes on vehicle front axles, studies and experimental investigations gradually decreased. According to Kusamo [32], the drum brake noise frequency increased with increasing brake hydraulic pressure; moreover, Lang [33] proposed the introduction of asymmetry into drum structure in order to reduce drum brake squeal. The frequency spectrum of drum brake noise is observed in the 500–4000 Hz range. Unlike brake noise, grabbing is a lower frequency vibration that is generally felt rather that heard, and is defined as a forced vibration. In order to find a solution to this friction-induced vibration and to minimize it, the effect of suspension and vehicle body dynamics on the transmission of grabbing to the driver have been investigated; the frequency spectrum of grabbing vibration is in the 10 – 100 Hz range.

## 3 ANALYTICAL MODEL

In a previous work [34], Boudot presented heavy trucks grabbing. According to experimental investigations, the dynamic characteristics of the whole front axle assembly is concerned, even if the source of grabbing is located in the braking system. The dynamic system is defined in Figure 2. We assume that the brake friction coefficient $\mu$ is constant.

Figure 2 : Dynamic model of braking system

Grabbing vibration results from coupling between the torsional mode $(k_2, m_2)$ of the front axle and the normal mode $(k_1, m_1)$ of the brake control. In order to simulate braking system placed crosswise due to overhanging cuased by static force effect, we may consider the moving belt slopes with an angle $\theta$. The braking force $F_{brake}$ transits through the braking command, that has non-linear behaviour. Therefore, we consider the possibility of having a non-linear contribution. Then, we can express this non-linear stiffness as a quadratic and cubic polynomial in the relative displacement:

$$\begin{aligned} k_1 &= k_{11} + k_{12}.(Y-y) + k_{13}.(Y-y)^2 \\ k_2 &= k_{21} + k_{22}.X + k_{23}.X^2 \end{aligned} \tag{1}$$

This nonlinearity is applied to indicate the influence and the importance of non-linear terms in the understanding of the dynamic behaviour of systems with non-linear phenomena . To be more precise, the non-linear dynamic behaviour of the front axle assembly and the non-linear dynamic behaviour of the brake command are concerned.

We assume that the tangential force $T$ is generated by the brake friction coefficient $\mu$, considering the Coulomb's friction law $T = \mu.N$ .

The three equations of motion can be expressed as:

$$\begin{cases} m_1\ddot{Y} + c_1(\dot{Y}-\dot{y}) + k_{11}(Y-y) + k_{12}(Y-y)^2 + k_{13}(Y-y)^3 = -F_{brake} \\ m_2\ddot{X} + c_2\dot{X} + k_{21}X + k_{22}X^2 + k_{23}X^3 = -N\sin\theta + T\cos\theta \\ m_2\ddot{y} + c_1(\dot{y}-\dot{Y}) + k_{11}(y-Y) + k_{12}(y-Y)^2 + k_{13}(y-Y)^3 = N\cos\theta + T\sin\theta \end{cases} \tag{2}$$

Using the transformations $y = X\tan\theta$ and $x = \{X \quad Y\}^T$, and considering the Coulomb's friction law $T = \mu.N$ , the non-linear 2-degrees-of-freedom system has the form

$$[M].\{\ddot{x}\}+[C].\{\dot{x}\}+[K].\{x\}=\{F\}+\{F_{nonlinear}\} \tag{3}$$

Where $\{\ddot{x}\}$, $\{\dot{x}\}$ and $\{x\}$ are the acceleration, velocity, and displacement response *2*-dimensional vector of the degrees-of-freedom, respectively; $[M]$ is the mass matrix, $[C]$ the damping matrix, and $[K]$ the stiffness matrix; $\{F\}$ is the vector force due to brake command and $\{F_{nonlinear}\}$ contains moreover the non-linear stiffness terms. The base parameters are defined in Appendix A. We have

$$[M]=\begin{bmatrix} m_2\left(\tan^2\theta+1\right) & 0 \\ 0 & m_1 \end{bmatrix} \tag{4}$$

$$[C]=\begin{bmatrix} c_1\left(\tan^2\theta-\mu\tan\theta\right)+c_2\left(1+\mu\tan\theta\right) & c_1\left(-\tan\theta+\mu\right) \\ -c_1\tan\theta & c_1 \end{bmatrix} \tag{5}$$

$$[K]=\begin{bmatrix} k_{21}\left(1+\mu\tan\theta\right)+k_{11}\left(\tan^2\theta-\mu\tan\theta\right) & k_{11}\left(-\tan\theta+\mu\right) \\ -k_{11}\tan\theta & k_{11} \end{bmatrix} \tag{6}$$

$$\{F_{nonlinear}\}=\begin{Bmatrix} \left(-\tan\theta+\mu\right)\left(k_{12}\left(X\tan\theta-Y\right)^2+k_{13}\left(X\tan\theta-Y\right)^3\right)+k_{22}\left(1+\mu\tan\theta\right)X^2+k_{23}\left(1+\mu\tan\theta\right)X^3 \\ -k_{12}\left(Y-X\tan\theta\right)^2-k_{13}\left(Y-X\tan\theta\right)^3 \end{Bmatrix} \tag{7}$$

$$\{F\}=\begin{Bmatrix} 0 \\ -F_{brake} \end{Bmatrix} \tag{8}$$

## 4 STABILITY AND HOPF BIFURCATION POINT

The first step is the static problem; the steady state operating point for the full set of non-linear equations is obtained by the determination of the equilibrium point. We obtain the linearized equations of motion by the introduction of small perturbations at the equilibrium point into the non-linear equations. Stability is investigated by the determination of eigenvalues of these linearized equations for each steady-state operating point of the non-linear system.

The equilibrium point $\{x_0\}$ is obtained by solving the non-linear static equations for a given net brake hydraulic pressure; this equilibrium point satisfies the following conditions:

$$K.\{x_0\}=\{F_{brake}\}+\{F_{nonlinear}(x_0)\} \tag{9}$$

Since the sprag-slip equations are non-linear, more than one steady-state operating point at a given net brake hydraulic pressure can be obtained.

The stability of the system is investigated on the linearized equations by assuming small perturbations $\{\overline{x}\}=\{\overline{X} \quad \overline{Y}\}^T$ at the equilibrium point $\{x_0\}=\{X_0 \quad Y_0\}^T$ of the non-linear system $\{x\}=\{x_0\}+\{\overline{x}\}$.

Substituting this expression into the non-linear equations (3), and neglecting higher order terms, we obtain the linearized equations of motion:

$$[M].\{\ddot{\overline{x}}\}+[C].\{\dot{\overline{x}}\}+[K].\{\overline{x}\}=\{F^L_{nonlinear}(\overline{x})\} \quad \text{with} \quad F^L_{nonlinear}(\overline{x})=\left.\frac{\partial F_{nonlinear}}{\partial \overline{x}_i}\right|_{x_0}.\overline{x}_i \tag{10}$$

Now, stability analyses can be performed on the linearized equations for small perturbations at the operating point of the non-linear systems. The eigenvalues of this system can be expressed $\lambda = a + i.b$. If $a$ is negative or equal to zero, the system is stable and we don't have vibration. If $a$ is positive, we have an unstable root and vibration. Moreover, $b$ represents frequency of the unstable mode. Computations are conducted with various brake friction coefficients. The Hopf bifurcation point is detected for $\mu_0 = 0,204$ and defined as follows:

$$\begin{cases} Re(\lambda(\mu))\big|_{x=x_0,\mu=\mu_0} = 0 \\ \dfrac{d}{d\mu}\left(Re[\lambda(\mu)]\right)\Big|_{\mu=\mu_0} \neq 0 \end{cases} \tag{11}$$

A representation of the evolution of frequencies and the associated real parts against brake friction coefficient are plotted in Figure 3 and in Figure 4, respectively. As illustrated in Figure 3, there are two stable modes at different frequencies when $\mu < \mu_0$. On the other hand, as illustrated in Figure 4, the real part of eigenvalues is negative when $\mu < \mu_0$. For $\mu = \mu_0$, there is one pair of purely imaginary eigenvalues and we have the Hopf bifurcation point. All other eigenvalues have negative real parts. After the bifurcation, the two modes couple and form a complex pair. On the other hand, the real part of eigenvalues is positive. As shown in Figure 4, the system is unstable for $\mu > \mu_0$ and stable for $\mu < \mu_0$. Moreover, the frequency $\omega_0$ of the unstable mode obtained for $\mu = \mu_0$ is near 50 Hz. There is a perfect correlation with experiment tests where grabbing vibration is in the 40-70 Hz range.
Therefore, it is possible to characterise the stability properties of the linearized system by representing the evolution of the eigenvalues by variation of $\mu$ in the complex plane, as illustrated in Figure 5.

Figure 3 : Coupling of two eigenvalues
Figure 4 : Evolution of the real part of two coupling modes
Figure 5 : Evolution of the eigenvalues by variation of $\mu$ in the complex plane

## 5 CENTER MANIFOLD APPROACH AND LIMIT CYCLE

In this section, we briefly described the center manifold approach in the proximity of a bifurcation point. The center manifold approach can be compared as a simplification method that reduces the number of equations of the original system (Nayfeh [8], Guckhenheimer [11] and Knobloch [18]).
Previously, stability analyses were investigated by determining eigenvalues of the linearized equations at each steady-state operating point for small perturbations. In order to carry out a complex non-linear analysis, it is necessary to consider the complete expressions of the non-linear forces. Moreover, the complete non-linear expressions are expressed at the equilibrium point for small perturbations. The non-linear sprag-slip equations at the equilibrium point $\{x_0\} = \{X_0 \quad Y_0\}^T$ for small perturbations $\{\overline{x}\} = \{\overline{X} \quad \overline{Y}\}^T$ can be expressed as

$$[M].\{\ddot{\overline{x}}\} + [C].\{\dot{\overline{x}}\} + [K].\{\overline{x}\} = \sum_{i=1}^{2} f_{(1)}^{i}.\overline{x}_i + \sum_{i=1}^{2}\sum_{j=1}^{2} f_{(2)}^{ij}.\overline{x}_i.\overline{x}_j + \sum_{i=1}^{2}\sum_{j=1}^{2}\sum_{k=1}^{2} f_{(3)}^{ijk}.\overline{x}_i.\overline{x}_j.\overline{x}_k \tag{12}$$

where the vectors $f_{(1)}^{i}$, $f_{(2)}^{ij}$ and $f_{(3)}^{ijk}$ are the coefficients of the linear, quadratic, and cubic terms, respectively , due to the non-linear stiffness at the equilibrium point. The expressions of $f_{(1)}^{i}$, $f_{(2)}^{ij}$ and $f_{(3)}^{ijk}$ are given in appendix B.

In order to use the center manifold approach, we write the non-linear equation in state variables $y = \{\bar{x} \quad \dot{\bar{x}}\}$

$$\dot{y} = f(y,\mu) = A(\mu).y + \sum_{i=1}^{4}\sum_{j=1}^{4}\eta_{(2)}^{ij}(\mu).y_i.y_j + \sum_{i=1}^{4}\sum_{j=1}^{4}\sum_{k=1}^{4}\eta_{(3)}^{ijk}(\mu).y_i.y_j.y_k \tag{13}$$

where $\mu$ is the friction parameter, $A(\mu)$, $\eta_{(2)}^{ij}$, and $\eta_{(3)}^{ijk}$ are the $4\times 4$ matrix, quadratic and cubic non-linear terms, respectively. We have

$$A = -\begin{bmatrix} C & M \\ I & 0 \end{bmatrix}^{-1} . \begin{bmatrix} \tilde{K} & 0 \\ 0 & I \end{bmatrix} \tag{14}$$

$$\eta_{(2)} = \begin{bmatrix} C & M \\ I & 0 \end{bmatrix}^{-1} . \begin{bmatrix} f_{(2)} \\ 0 \end{bmatrix} \tag{15}$$

$$\eta_{(3)} = \begin{bmatrix} C & M \\ I & 0 \end{bmatrix}^{-1} . \begin{bmatrix} f_{(3)} \\ 0 \end{bmatrix} \tag{16}$$

The problem can be put into Jordan normal form by means of the eigenbasis. At the Hopf bifurcation point, the previous system can be written in the form

$$\begin{cases} \dot{v}_c = J_c.v_c + G_2(v_c,v_s) + G_3(v_c,v_s) \\ \dot{v}_s = J_s.v_s + H_2(v_c,v_s) + H_3(v_c,v_s) \end{cases} \tag{17}$$

where $J_c$ and $J_s$ have eigenvalues $\lambda$ such as $Re[\lambda_{Jc}(\mu_0)] = 0$ and $Re[\lambda_{Js}(\mu_0)] \neq 0$. $G_2, G_3, H_2$ and $H_3$ are polynomials of degree 2 and 3 versus $v_c$ and $v_s$. Since, we consider here the physically interesting case of the stable equilibrium losing stability; we assume that the unstable manifold is empty. The center manifold theory enables us to express the variables $v_s$ as a function of $v_c$: $v_s = h(v_c)$ (Carr [35], 1981). The expression of $h$ cannot be solved explicitly. However, it is possible to define an approximate solution of $h$ by a power expansion and by equating the coefficients. In order to satisfy the tangency conditions at the bifurcation point to the center eigenspace, the function $h$ verifies $h(0)=0$ and $Dh(0)=0$. So, we define $v_s = h(v_c)$ as a power series in $v_c$ , without constant and linear terms. By differentiating and substituting the center manifold $v_s = h(v_c)$ into the second equation of Eq.(17), we obtain

$$D_{v_c}(h(v_c)).(J_c.v_c + G_2[v_c,h(v_c)] + G_3[v_c,h(v_c)]) = J_s.h(v_c) + H_2[v_c,h(v_c)] + H_3[v_c,h(v_c)] \tag{18}$$

A system of algebraic equations for the coefficients of the polynomials is obtained by equating the coefficients of the different terms in the polynomials on both sides. Solving these equations gives a first approximation of the center manifold $v_s = h(v_c)$. Now, the limit cycles are obtained for parameter values near the bifurcation point $\mu = \mu_0 + \bar{\mu}$ where $\mu_0$ is the bifurcation point and $\bar{\mu} = \varepsilon.\mu_0$ (with $\varepsilon$ <<1). We use an application of the center manifold approach by augmenting the system with the equation $\dot{\mu} = 0$. We assume that $\mu = O(v_c^2)$, so the local center manifold is represented by the polynomial expansion $v_s = h(v_c)$ as defined previously. This method is a simple extension to the center manifold method which is useful when dealing with parameterised families of systems. Moreover, the non-linear terms are approximated by their evaluations at the bifurcation point $\mu = \mu_0$; effectively since the unfolding parameter $\bar{\mu} = \varepsilon.\mu_0$ is very small (with $\varepsilon$ <<1), so the approximation $G_2[v_c,h(v_c),\mu_0]$ and $G_3[v_c,h(v_c),\mu_0]$ are equivalent to $G_2[v_c,h(v_c),\mu]$ and $G_3[v_c,h(v_c),\mu]$ with negligible error.

Finally, the dynamics are given by

$$\begin{cases} \dot{v}_c = J_c(\mu).v_c + G_2[v_c, h(v_c), \mu_0] + G_3[v_c, h(v_c), \mu_0] \\ \dot{\mu} = 0 \end{cases} \tag{19}$$

This reduced system is easier to study than the original system. . Using an approximation of $h$ of order 2 causes divergence in the evolutions of limit cycles. Using approximations of $h$ of order 3 or 4 allow to obtain a first approximation of the limit cycles but we consider that it is not sufficient, as illustrated in Figure 6 and in Figure 7. Using an approximation of $h$ of order 5 allows to obtain a very good estimation of the limit cycles as illustrated in Figure 6 and in Figure 7.

Figure 6 : X-limit cycle for $\overline{\mu} = 1/1000.\mu_0$

Figure 7 : Y-limit cycle for $\overline{\mu} = 1/1000.\mu_0$

## 6 SYSTEM'S TRANSFORMATION TO USE MULTIVARIABLES APPROXIMANTS

Our purpose in this section is to write again the system defined in (19) in terms of $v_c = \{v_{c1} \quad v_{c2}\}^T$ in order to use the multivariable approximants for the simplification of the non-linear system. Considering $v_s = h(v_c)$ as a power series in $v_c$ of degree $m$ without constant and linear terms ($m \geq 2$), we note that $G_2[v_c, h(v_c), \mu_0]$ and $G_3[v_c, h(v_c), \mu_0]$ are power series in $v_c$ of degree $2m$ and $3m$, without constant and linear terms ($m \geq 2$). We have :

$$\begin{cases} G_2[v_c, h(v_c)] = \begin{bmatrix} g_{2.11} & \cdots & g_{2.116} \\ g_{2.12} & \cdots & g_{2.216} \end{bmatrix} . \begin{Bmatrix} v_{c1} \\ v_{c2} \\ h_1(v_{c1}, v_{c2}) \\ h_2(v_{c1}, v_{c2}) \end{Bmatrix} \otimes \begin{Bmatrix} v_{c1} \\ v_{c2} \\ h_1(v_{c1}, v_{c2}) \\ h_2(v_{c1}, v_{c2}) \end{Bmatrix} = \begin{Bmatrix} \sum_{p=2}^{2m} \sum_{i=0}^{p} \varphi_{1.i,p-i} . v_{c1}^i . v_{c2}^{p-i} \\ \sum_{p=2}^{2m} \sum_{i=0}^{p} \varphi_{2.i,p-i} . v_{c1}^i . v_{c2}^{j} \end{Bmatrix} = g_2(v_c) \\ G_3[v_c, h(v_c)] = \begin{bmatrix} g_{3.11} & \cdots & g_{3.164} \\ g_{3.12} & \cdots & g_{3.264} \end{bmatrix} . \begin{Bmatrix} v_{c1} \\ v_{c2} \\ h_1(v_{c1}, v_{c2}) \\ h_2(v_{c1}, v_{c2}) \end{Bmatrix} \otimes \begin{Bmatrix} v_{c1} \\ v_{c2} \\ h_1(v_{c1}, v_{c2}) \\ h_2(v_{c1}, v_{c2}) \end{Bmatrix} \otimes \begin{Bmatrix} v_{c1} \\ v_{c2} \\ h_1(v_{c1}, v_{c2}) \\ h_2(v_{c1}, v_{c2}) \end{Bmatrix} = \begin{Bmatrix} \sum_{p=2}^{3m} \sum_{i=0}^{p} \gamma_{1.i,p-i} . v_{c1}^i . v_{c2}^{p-i} \\ \sum_{p=2}^{3m} \sum_{i=0}^{p} \gamma_{2.i,p-i} . v_{c1}^i . v_{c2}^{p-i} \end{Bmatrix} = g_3(v_c) \end{cases} \tag{20}$$

where $\varphi_{i,p-i}$ and $\gamma_{i,p-i}$ are matrices of the constant coefficients and $\otimes$ defines the Kronecker product $\otimes$(Stewart [36]).

The substitution of Eq.(20) in Eq.(19) gives

$$\begin{Bmatrix} \dot{v}_{c1} \\ \dot{v}_{c2} \end{Bmatrix} = [J_c(\mu)] . \begin{Bmatrix} v_{c1} \\ v_{c2} \end{Bmatrix} + \begin{Bmatrix} \sum_{p=2}^{2m} \sum_{i=0}^{p} \varphi_{1.i,p-i} . v_{c1}^i . v_{c2}^{p-i} \\ \sum_{p=2}^{2m} \sum_{i=0}^{p} \varphi_{2.i,p-i} . v_{c1}^i . v_{c2}^{p-i} \end{Bmatrix} + \begin{Bmatrix} \sum_{p=2}^{3m} \sum_{i=0}^{p} \gamma_{1.i,p-i} . v_{c1}^i . v_{c2}^{p-i} \\ \sum_{p=2}^{3m} \sum_{i=0}^{p} \gamma_{2.i,p-i} . v_{c1}^i . v_{c2}^{p-i} \end{Bmatrix} \tag{21}$$

where $[J_c(\mu)]$ is a $2 \times 2$ matrix containing the coefficients of linear terms. Finally, the system can be written as follow :

$$\begin{Bmatrix} \dot{v}_{c1} \\ \dot{v}_{c2} \end{Bmatrix} = \begin{Bmatrix} \sum_{p=1}^{3m} \sum_{i=0}^{p} c_{1.i,p-i} . v_{c1}^{i} . v_{c2}^{p-i} \\ \sum_{p=1}^{3m} \sum_{i=0}^{p} c_{2.i,p-i} . v_{c1}^{i} . v_{c2}^{p-i} \end{Bmatrix} \tag{22}$$

where $c_{1.ij}$ and $c_{2.ij}$ are the coefficients in terms of $v_{c1}^{i}.v_{c2}^{j}$ $(i+j \geq 1)$. Equating the coefficients of different terms in the polynomials on both sides, gives a system of algebraic equations for the determination of the coefficients $c_{1.ij}$ and $c_{2.ij}$. We note that the determination of $c_{1.ij}$ and $c_{2.ij}$ are completely independent. Moreover, we remark that $c_{1.10}$, $c_{1.01}$, $c_{2.10}$ and $c_{2.01}$ are only defined by the expression of $[J_c(\mu)]$ (we have $c_{1.10} = J_c(1,1)$, $c_{1.01} = J_c(1,2)$, $c_{2.10} = J_c(2,1)$ and $c_{2.01} = J_c(2,2)$). Hence, $c_{1.10}$, $c_{1.01}$, $c_{2.10}$ and $c_{2.01}$ are defined for $\mu = \mu_0 + \overline{\mu}$. In the other hand, $c_{1.ij}$ and $c_{2.ij}$ (with $i+j \geq 2$) are defined for $\mu_0$. Therefore, the system can be expressed as follows:

$$\begin{Bmatrix} \dot{v}_{c1} \\ \dot{v}_{c2} \end{Bmatrix} = \begin{Bmatrix} f_1(v_{c1}, v_{c2}) \\ f_2(v_{c1}, v_{c2}) \end{Bmatrix} = \begin{Bmatrix} c_{1.10}(\mu).v_{c1} + c_{1.01}(\mu).v_{c2} \\ c_{2.10}(\mu).v_{c1} + c_{2.01}(\mu).v_{c2} \end{Bmatrix} + \begin{Bmatrix} \sum_{p=2}^{3m} \sum_{i=0}^{p} c_{1.i,p-i}(\mu_0).v_{c1}^{i}.v_{c2}^{p-i} \\ \sum_{p=2}^{3m} \sum_{i=0}^{p} c_{2.i,p-i}(\mu_0).v_{c1}^{i}.v_{c2}^{p-i} \end{Bmatrix} \tag{23}$$

## 7 PADE APPROXIMANTS AND MULTIVARIABLE APPROXIMANTS

Let $f$ be a formal power series $f(z) = \sum_{i=0}^{\infty} c_i z^i$. Let us consider a rational function with a numerator of the degree $p$ at most a denominator of the degree $q$ at most such that its power series expansion (obtained by dividing the numerator and by the denominator in ascending powers of the variable $z$) agrees with that of $f$ as far as possible. Such a rational function is called an Padé approximant of $f$ and it is usually denoted $[p/q]_{f(z)}$. Its numerator has $p+1$ coefficients and its denominator $q+1$. But, since a rational function is defined apart from a multiplying factor, which will be taken so that the constant term of the denominator is equal to $1$, there are only $p+q+1$ unknown coefficients in this Padé approximant. The two series will agree at least up to the term of degree $p+q$ inclusively: we have $f(z) - [p/q]_{f(z)} = O\left(z^{p+q+1}\right)$.

Here, it is not our purpose to give a full account of Padé approximation that can be found in Baker's [19] and Brezinski's [38]-[20] works. We confine our attention to the generalisation of Padé approximants and more particularly we consider two-variable rational approximants. The main purpose of Padé approximants is to approximate functions given by a formal series expansion. We use the applications of the multivariable approximants in order to simplify the system. Moreover, the use of the approximants allows to obtain limit cycles more easily and rapidly.

We suppose that $\sum_{i=0}^{\infty} \sum_{j=0}^{\infty} c_{ij} x^i y^j$ is a power series representing a function $f(x,y)$. We define polynomials $N(x,y) = \sum_{i,j \in A} n_{ij} x^i y^j$ and $D(x,y) = \sum_{i,j \in B} d_{ij} x^i y^j$ so that

$$f(x,y) = \frac{N(x,y)}{D(x,y)} + \sum_{i=0}^{\infty} \sum_{j=0}^{\infty} e_{ij} x^i y^j \tag{24}$$

where as many coefficients $e_{ij}$ as possible are equal to zero. $A$ and $B$ define the lattice spaces (Baker [19]), for $N(x,y)$ and $D(x,y)$, respectively. The numerator and denominator coefficients are in lattice spaces $A$ and $B$. We require that $e_{ij}=0$ for $i,j\in E$, the equality space. Considering $b_{00}=1$ as part of the definition, this scheme is normally determinate if $dim(E)=dim(A)+dim(B)-1$.

The developments are known as the generalized Chisholm approximants or the Canterbury approximants. The case of Canterbury approximants we are considering is the general system of Hughes Jones approximants (Hughes Jones[39]), defined as

$$N^{[L/M]}(x,y)=\sum_{i=0}^{L}\sum_{j=0}^{L}n_{ij}x^i y^j \quad \text{and} \quad D^{[L/M]}(x,y)=\sum_{i=0}^{M}\sum_{j=0}^{M}d_{ij}x^i y^j \tag{25}$$

Returning to (24) and cross-multiplying, we find that

$$\begin{aligned}&\left(d_{00}+d_{10}x+d_{01}y+d_{11}xy+...+d_{MM}x^M y^M\right)\left(c_{00}+c_{10}x+c_{01}y+c_{11}xy+...\right)=\\ &\qquad n_{00}+n_{10}x+n_{01}y+n_{11}xy+...+n_{LL}x^L y^L+\sum_{i=0}^{\infty}\sum_{j=0}^{\infty}e_{ij}x^i y^j\end{aligned} \tag{26}$$

Then we have equality at order $x^\alpha y^\beta$ if

$$\sum_{i=0}^{\alpha}\sum_{j=0}^{\beta}d_{ij}c_{\alpha-i,\beta-j}=n_{\alpha\beta} \quad \text{for} \quad (\alpha,\beta)\in A \tag{27}$$

$$\sum_{i=0}^{min(\alpha,M)}\sum_{j=0}^{min(\beta,M)}d_{ij}c_{\alpha-i,\beta-j}=0 \quad \text{for} \quad (\alpha,\beta)\in E,(\alpha,\beta)\notin A \tag{28}$$

The numerator coefficients $n_{ij}$ are determined by (27), once the denominator coefficients $d_{ij}$ are determined. This method, called the prong method, is more detailed by Hughes Jones and Makinson [40] .

Using the multivariable approximants, we can rewrite the system defined in (23) as follows :

$$\begin{Bmatrix}\dot{v}_{c1}\\ \dot{v}_{c2}\end{Bmatrix}=\begin{Bmatrix}p_1(v_{c1}.v_{c2})\\ p_2(v_{c1}.v_{c2})\end{Bmatrix}=\begin{Bmatrix}\sum_{i=0}^{L}\sum_{j=0}^{L}n_{1.ij}.v_{c1}^i.v_{c2}^j \Big/ \sum_{i=0}^{M}\sum_{j=0}^{M}d_{1.ij}.v_{c1}^i.v_{c2}^j\\ \sum_{i=0}^{L}\sum_{j=0}^{L}n_{2.ij}.v_{c1}^i.v_{c2}^j \Big/ \sum_{i=0}^{M}\sum_{j=0}^{M}d_{2.ij}.v_{c1}^i.v_{c2}^j\end{Bmatrix} \tag{29}$$

where $p_1(v_{c1}.v_{c2})$ and $p_2(v_{c1}.v_{c2})$ are rational functions with a numerator of degree $L$ and a denominator of degree $M$ defined by using the multivariable approximants. The determination of the coefficients $n_{1.ij}$, $n_{2.ij}$ (for $i,j=0,1,..,L$) and $d_{1.ij}$, $d_{2.ij}$ (for $i,j=0,1,..,M$) is obtained by using the procedure defined previously. We note that the determination of $(n_{1.ij},d_{1.ij})$ and $(n_{2.ij},d_{2.ij})$ are completely independent. The use of multivariable approximants allows the computation of an accurate approximation of $f(v_{c1},v_{c2})$ even at values of $f$ for which the Taylor series of $f(v_{c1},v_{c2})$ diverge.

Applications of the multivariable approximants is applied in order to simplify the non-linear expression of Eq.(23), that is a power series in $v_c$ of degree $15$, without constant terms. In order to approximate the solution, we use $[5/4]_{f_1(v_{c1},v_{c2})}$ approximants and $[5/4]_{f_2(v_{c1},v_{c2})}$ approximants. We notice that an $[L/M]_{f(v_{c1},v_{c2})}$ approximation with $L\leq 4$ and $M\leq 4$ is not sufficient : effectively, in some cases, computations diverge since the non-linearities retained are not sufficient, and in other cases, the limit cycle amplitudes obtained are not acceptable due to the same reasons. An $[L/M]_{f(v_{c1},v_{c2})}$ approximation with $L \geq 5$ and $N \geq 4$ gives a good correlation with the original non-linear system. A representation of the limit cycle amplitudes for different $[L/M]_f$ approximants are plotted in Figure 8 and in Figure 9.

The interest of multivariable approximants is that they need less terms than the Taylor series in order to obtain an accurate approximation of the limit cycle amplitudes. Effectively, we need to use a center manifold at least of order 5 in order to have the same estimation of the limit cycles. So, the non-linear terms becomes a power series of degree 15 in which all terms are relevant.

Moreover, the determination of limit cycle amplitude by the integration of the differential-algebraic equations of the system is faster using the multivariable approximants.

Figure 8 : X-limit cycle for $\bar{\mu} = \mu_0 / 1000$

Figure 9 : Y-limit cycle for $\bar{\mu} = \mu_0 / 1000$

## 8 APPLICATION OF THE HARMONIC BALANCE

One of the systematic way for analysing non-linear systems is the harmonic balance method (HBM) in order to discretize the unknowns functions in time by their Fourier components, which are assumed to be constant with respect to time. Such methods include the incremental harmonic balance (IHB) method, the Fast Galerkin (FG) method, and the alternate frequency/time domain (AFT) method. Cheung, Chen and Lau [2], Leung and Chui [3], Lau and Zhang [4], Pierre, Ferri and Dowell [5] employed the incremental harmonic balance (IHB) method for different non-linear systems. The developments of procedures made by Ling and Wu [41] for the Fast Galerkin (FG) method and by Cameron and Griffin [6] for the alternate frequency/time domain (AFT) method essentially consist in a substitution of the harmonic balance step by a FFT algorithm.

In our case, we use the alternate frequency/time domain (AFT) method in order to obtain limit cycle amplitudes of the non-linear system defined in Eq.(29). In this case, we will show that it is possible to carry out the AFT method and to obtain the correct limit cycle amplitudes in the form of a Fourier series by considering the non-linear simplified system with multivariable approximants form. We consider the general multi-degree-of-freedom non-linear autonomous dynamical system corresponding to Eq.(29) in the form

$$\dot{z}_i + f_i(z) = 0 \quad , \quad i = 1,...,N \tag{30}$$

where $f_i$ represents the non-linear expression in multivariable approximants form, as described previously, and defined by $p_i(v_{c1}, v_{c2})$ . We remark that this system is an autonomous system. The frequency of oscillation is known, since we will obtained a limit cycle amplitude near the Hopf bifurcation point $\mu = \mu_0 + \bar{\mu}$ where $\mu_0$ is the bifurcation point and $\bar{\mu} = \varepsilon.\mu_0$ (with $\varepsilon$ <<1). So, the exact frequency is obtained by the determination of the equilibrium point and eigenvalues of these linearized equations for $\mu = \mu_0 + \bar{\mu}$, as defined previously in Eq.(9) and Eq.(10), respectively. The frequency is equal to 50.2 Hz, so $\omega = 316 rad/s$ . We considering the truncated Fourier series expansion for $z_i^k$ :

$$z_i^k = Z_{i,0}^k + \sum_{j=1}^{M} \left[ Z_{i,2j-1}^k . cos(j\omega t) + Z_{i,2j}^k . sin(j\omega t) \right] \qquad \text{for } i = 1,..,N \tag{31}$$

where $Z_{i,0}^{k}$, $Z_{i,2j-1}^{k}$, $Z_{i,2j}^{k}$ are the Fourier coefficients. The number of harmonic coefficients $M$ is selected in order to consider only the significant harmonics expected in the solution. We obtain $(2M+1)N$ linear algebraic equations

$$[A].\{Z^{k+1}\}+\{F^{NL}\}+[A+J].\{\Delta Z^{k}\}=\{0\} \tag{32}$$

where $[A]$ and $[J]$ are the Jacobian matrices associated with the linear and non-linear parts of the equation of motion (30). They are defined in Appendix C. $\{F^{NL}\}$ represents the vector of the Fourier coefficients of the non-linear function $f$. $\{Z^{k+1}\}$ and $\{\Delta Z^{k}\}$ contain the Fourier coefficients and the Fourier increments of $z^{k}$ and $\Delta z^{k}$, respectively. Equation (32) contains $(2M+1)N$ unknowns which can be determined by solving the $(2M+1)N$ linear algebraic equations. $\{F^{NL}\}$ is difficult to determine from the Fourier coefficients directly for many non-linear elements. Hence Cameroun and Griffin [6] suggested to calculate $\{F^{NL}\}$ following a path (alternating frequency/time domain AFT) as follows

$$Z \xrightarrow{DFT^{-1}} z(t) \Rightarrow f(t) \xrightarrow{DFT} F^{NL} \tag{33}$$

We use the Discrete Fourier Transform (DFT) as defined by Narayanan and Sekar [7]. It is given in Appendix C. The number of collocation nodes are calculated as $(2M+1)N$, where $M$ is the number of harmonics considered in the Fourier series expansions.
The error vector $\{R\}$ is given by

$$\{R\}=[A].\{Z^{k+1}\}+\{F^{NL}\} \tag{34}$$

and the convergence is chosen and given by

$$\delta=\sqrt{R_0^2+\sum_{j=1}^{M}\left(R_{2j-1}^2+R_{2j}^2\right)} \tag{35}$$

An initial estimate of the Fourier coefficients is necessary in order to start the iteration scheme. The complete scheme of the computer program using the alternate frequency/time domain (AFT) method with the Discrete Fourier Transform (DFT) is expressed in Figure 10.

Figure 10 : Direct Iteration of the DFT method

Computations are conducted by using various power harmonics. We note that one harmonic ($M=1$) is not sufficient to describe exactly the limit cycle amplitude. Conversely, the second or more order of harmonic coefficients ($M=2$, $M=3$ or $M\geq 3$) allow the obtention of the same limit cycle amplitudes than that obtained by the integration of the original system. The values of the Harmonic coefficients for one and two harmonics are given in Table 1. Moreover, we note that the values of the harmonic coefficients are complex, since they defined the unknown functions in time of $(v_{c1}, v_{c2})$ by their Fourier components in the center manifold base.
Using the reverse transformation in order to go from the center manifold space (with complex variable) to the physical space (with real variable), we obtain the limit cycle amplitude of the non-linear system defined in Eq.(3) near the bifurcation point $\mu=\mu_0+\bar{\mu}$, where $\mu_0$ is the bifurcation point and $\bar{\mu}=\varepsilon.\mu_0$ (with $\varepsilon$ <<1). A representation of the limit cycle amplitudes for different orders are plotted in Figure 11 and in Figure 12.We note that this method is ideal for parametric studies since we can go from a known state of vibration to the next neighboring state which corresponds to an incremental change in one of the parameters of the system. Therefore, the method is rapidly convergent.

Table 1 : Values of the harmonic coefficients
Figure 11 : X-limit cycles amplitude by using the alternate frequency/time domain method
Figure 12 : Y-limit cycles amplitude by using the alternate frequency/time domain method

## 9 SUMMARY AND CONCLUSION

A non-linear model for the analysis of mode heavy truck grabbing has been developed. Results from stability have been investigated by determining eigenvalues of this linearized equation for each steady-state operating point of the non-linear system. This stability analysis indicates that system instability can occur with a constant friction coefficient.

Moreover, this paper presents a procedure using successively the center manifold approach, the multivariable approximants, and the AFT method in order to obtain equations for the limit cycle amplitude. This approach simplifies the dynamics on the centre manifold by reducing the order of the dynamical system, while retaining the essential features of the dynamic behaviour near the Hop bifurcation point. The multivariable approximants allow the approximation of the non-linear system as rational non-linear equations. One of the most important points is the possibility to use successfully these two methods in order to obtain a reduced and simplified non-linear system, without losing contributions of non-linear terms. Finally, the AFT method is applied to discretize the unknown functions in time by their Fourier components. We have validate this procedure by comparing the results obtained by solving the full system, and by using the combination of center manifold approach, multivariable approximants, and AFT method. This approach seems very interesting when time history response solutions of the full set of non-linear equations are both time consuming and costly. Moreover, extensive parametric design studies can be done in order to relate the effect of specific parameter variation on the stability and the evolution of limit cycle amplitude.

## ACKNOWLEDGMENTS

The authors gratefully acknowledge the French Education Ministry for its support through grant n°99071 for the investigation presented here.

## REFERENCE

[1] A.H. Nayfeh and D.T. Mook, *Nonlinear Oscillations*. New-York : John Wiley & Sons, 1979.

[2] Y.K. Cheung, S.H. Chen and S.L. Lau, *Application of the incremental Harmonic Balance Method to Cubic Non-linearity Systems*, Journal of Sound and Vibration, 140(2), p273-286, 1990.

[3] A.Y.T Leung and S.K. Chui, *Non-linear Vibration of Coupled Duffing Oscillators by an Improved Incremental Harmonic Balance Method*, Journal of Sound and Vibration, 181(4), p619-633, 1995.

[4] S.L. Lau and W.S. Zhang, *Nonlinear Vibrations of piecewise-linear Systems by Incremental Harmonic Balance Method*, Journal of Applied Mechanics, 59, p153-160, 1992.

[5] C. Pierre, A.A. Ferri and E.H. Dowell, *Multi-Harmonic analysis of dry friction damped systems using an incremental harmonic balance method*, Journal of Applied Mechanics, 52, p958-964, 1985.

[6] T.M. Cameron and J.H. Griffin, *An Alternating Frequency/time Domain method for calculating the steady state response of nonlinear dynamic,* Journal of Applied Mechanics, 56, p149-154, 1989.

[7] S. Narayanan and P. Sekar, *A Frequency Domain Based Numeric-Analytical Method for Non-linear Dynamical Systems*, Journal of Sound and Vibration, 211(3), p409-424, 1998.

[8] A.H. Nayfeh and B. Balachandran, *Applied Nonlinear Dynamics : Analytical, Comptational and Experimental Methods*, John Wiley & Sons, 1995.

[9] A.D. Brjuno, *Transactions of the Moscow Mathematical Society 25, Analytical forms of differential equations, I*, p132-198, 1971.

[10] A.D. Brjuno, Transactions of the Moscow Mathematical Society 25, *Analytical forms of differential equations, II*, p199-299, 1972.

[11] J. Guckenheimer and P. Holmes, *Nonlinear Oscillations, Dynamical Systems, and Bifurcations of vector Fields,* Springer-Verlag, 1986.

[12] L. Jezequel and C.H. Lamarque, *Analysis of Non-linear Dynamical Systems by the Normal Form Theory*. Journal of Sound and Vibration, 149, p429-459,1991.

[13] C. Elphick, E. Tiraipegui, M.E. Brochet, P. Couillet, G. Iooss, *A simple global Characterization for Normal Forms of Singular vector Fields*, Phys.D, Preprint n°109, Université de Nice, 1986.

[14] G. Iooss. and D.D. Joseph, *Elementary Bifurcation and Stability Theory*. Berlin : Springer-Verlag, 1980.

[15] L. Hsu, *Analysis of critical and post-critical behaviour of non-linear dynamical systems by the normal form method, part I : Normalisation formulae*. Journal of Sound and Vibration, 89, p169-181, 1983.

[16] L. Hsu, *Analysis of critical and post-critical behaviour of non-linear dynamical systems by the normal form method, part II : Divergence and flutter*. Journal of Sound and Vibration, 89, p183-194, 1983.

[17] J.E. Marsden and M. McCracken, *The Hopf Bifurcation and its Applications*. New-York: Spring-Verlag, Applied Mathematical Sciences 19, 1976.

[18] E. Knobloch and K.A. Wiesenfeld, *Bifurcation in Fluctuating Systems : The Center Manifold Approach*, Journal of Statistical Physics, 33, No 3, p611-637, 1983.

[19] G.A. Baker, P. Graves-Morris, *Padé Approximants*, Cambridge University press, Cambridge, 1996.

[20] C. Brezinski, *Padé Type Approximation and General Orthogonal Polynomials*, INSNM, vol. 50, Birkhauser-Verlag, Basel, 1980.

[21] P. Chambrette, *Stabilité des Systèmes Dynamiques avec Frottement sec : Application au Crissement des freins à disque,* Thèse de Doctorat, Ecole Centrale de Lyon, 1991.

[22] M.R. North, *A Mechanism of disc brake squeal* , 14th FISITA congress paper 1/9 ,1972.

[23] S.Y. Liu, M.A. Ozbek and J.T. Gordon, *A Nonlinear Model for Aircraft Brake Squeal Analysis. Part i : Model Description and Solution Methodology,* In ASME Design Engineering Technical Conferences, 3, 1996.

[24] M.J. Rudd, *Wheel/Rail Noise – Part II: Wheel Squeal*, Journal of Sound and Vibration, 46, No 3, p381-394, 1976.

[25] R.P. Jarvis and B. Mills. *Vibrations induced by dry friction*, Proc. Instn. Mech. Engrs. Vol 178, n°32, p847-866, 1963/1964.

[26] S.W.E Earles and C.K. Lee, *Instabilites arising from the frictional interaction of a pin-disc system resulting in noise generation*. Trans ASME, J. Engng Ind., 98, Series B, n°1, p81-86, 1976.
[27] R.T. Spurr, *A theory of brake squeal*, Proc. Auto. Div., Instn. Mech. Engrs, n°1 , p33-40, 1961-62.
[28] R.A. Ibrahim, *Friction-Induced Vibration, Chatter, Squeal and Chaos : Part I - Mechanics of Contact and Friction*, ASME Applied Mechanics Review, 47, No 7, p209-226, 1994.
[29] R.A. Ibrahim, *Friction-Induced Vibration, Chatter, Squeal and Chaos : Part II – Dynamics and Modeling*, ASME Applied Mechanics Review, 47, No 7, p227-253, 1994.
[30] J.T. Oden and J.A.C. Martins, *Models and Computational Methods for Dynamic friction Phenomena*, Computer Methods in Apllied Mechanics and Engineering, 52, p527-634, 1985.
[31] D.A. Crolla and A.M. Lang, *Brake Noise and Vibration – State of Art*, Tribologie Series 18 – Vehicle Tribology – Paper VII, p165-174 – Dowson, Taylor & Godet éditeurs, Elsevier.
[32] M. Kusano, H. Ishidou, S. Matsumura and S. Washizu *SAE*, Paper 851465. Experimental Study on the Reduction of drum Brake Noise.
[33] A.M. Lang and T.P. Newcomb, Paper C382/051, I.*Mech.E/EAEC Conf.*, Strasbourg, Paper C382/051, An experimental investigation into drum brake squeal, 1989.
[34] J.P. Boudot, *Modélisation des bruits de freinage des véhicules Industriels*, Thèse de Doctorat, Ecole Centrale de Lyon, 1995.
[35] J. Carr, *Application of Center Manifold*, Springer-Verlag, New-York, 1981.
[36] G.W. Stewart and Ji-guang Sun, *Computer Science and Scientific Computing. Matrix Perturbation Theory*, Academic Press, 1990.
[37] C. Brezinski, *Extrapolation algorithms and Padé approximations: a historical survey*, Applied Numerical Mathematics, 20, p299-318, 1983.
[38] C. Brezinski, *An introduction to Padé approximations*, in Curves and Surfaces in Geometric Design, P.J. Laurent, A. Le Méhauté, L.LL Schumaker eds., AK. Peters, Wellesley, p59-65, 1994.
[39] R. Hughes Jones, *General rational approximants in N-variables*, J. Approx. Theory, 16, p201-33 [7§6], 1976.
[40] R. Hughes Jones and G.J. Makinson, *The generation of Chisholm rational approximants to power series in two variables*, J. Inst. math. Appl., 13, p299-310, 1974.
[41] F.H. Ling and X.X. Wu , *Fast Galerkin method and its application to determine periodic solutions of non-linear oscillators*, International Journal of Non-Linear Mechanics, 22, p89-98, 1987.

## APPENDIX A : PARAMETER VALUES

| | |
|---|---|
| $F_{brake} = 1N$ | brake force |
| $m_1 = 1kg$ | equivalent mass of first mode |
| $m_2 = 1kg$ | equivalent mass of second mode |
| $c_1 = 5N/m/\sec$ | equivalent damping of first mode |
| $c_2 = 5N/m/\sec$ | equivalent damping of second mode |
| $k_{11} = 1.10^5 N/m$ | coefficient of linear term of stiffness $k_1$ |
| $k_{12} = 1.10^6 N/m^2$ | coefficient of quadratic term of stiffness $k_1$ |
| $k_{13} = 1.10^6 N/m^3$ | coefficient of cubic term of stiffness $k_1$ |
| $k_{21} = 1.10^5 N/m$ | coefficient of linear term of stiffness $k_2$ |
| $k_{22} = 1.10^5 N/m^2$ | coefficient of quadratic term of stiffness $k_2$ |
| $k_{23} = 1.10^5 N/m^3$ | coefficient of cubic term of stiffness $k_2$ |
| $\theta = 0,2 rad$ | sprag-slip angle |
| $\mu_0 = 0,204$ | brake friction coefficient at the Hopf bifurcation point |

## APPENDIX B : MATRICES OF THE SYSTEM AND DEFINITION OF $f_{(1)}^{i}$ , $f_{(2)}^{ij}$ AND $f_{(3)}^{ijk}$ COEFFICIENTS

The vectors $f_{(1)}^{i}$, $f_{(2)}^{ij}$ and $f_{(3)}^{ijk}$ are the coefficients of the linear, quadratic, and cubic terms of the nonlinear force $\{F_{nonlinear}\}^T$ , respectively , due to the nonlinear stiffness at the equilibrium point. The non-zero components of the vectors $\{f_{(1)}^{i}\} = \{f_{(1)}^{X,i}\ f_{(1)}^{Y,i}\}^T$ , $\{f_{(2)}^{ij}\} = \{f_{(2)}^{X,ij}\ f_{(2)}^{Y,ij}\}^T$ and $\{f_{(3)}^{ij}\} = \{f_{(3)}^{X,ijk}\ f_{(3)}^{Y,ijk}\}^T$ ,respectively , are :

$$f_{(1)}^{X,1} = (-\tan\theta + \mu)\left[2k_{12}\tan^2\theta.X_0 - 2k_{12}\tan\theta.Y_0 + 3k_{13}\tan^3\theta.X_0^{\ 2} - 6k_{13}\tan^2\theta.X_0.Y_0 + 3k_{13}\tan\theta.Y_0^{\ 2}\right]$$
$$+(1+\mu\tan\theta)\left[2k_{22}.X_0 + 3k_{23}.X_0^{\ 2}\right]$$
$$f_{(1)}^{X,2} = (-\tan\theta + \mu)\left[2k_{12}.Y_0 - 2k_{12}\tan\theta.X_0 - 3k_{13}\tan^2\theta.X_0^{\ 2} + 6k_{13}\tan\theta.X_0.Y_0 - 3k_{13}.Y_0^{\ 2}\right]$$

$$f_{(1)}^{Y,1} = -2k_{12}\tan^2\theta.X_0 + 2k_{12}\tan\theta.Y_0 + 3k_{13}\tan\theta.Y_0^{\ 2} - 6k_{13}\tan^2\theta.X_0.Y + 3k_{13}\tan^3\theta.X_0^{\ 2}$$
$$f_{(1)}^{Y,2} = -2k_{12}.Y_0 + 2k_{12}\tan\theta.X_0 - 3k_{13}.Y_0^{\ 2} + 6k_{13}\tan\theta.X_0.Y_0 - 3k_{13}\tan^2\theta.X_0^{\ 2}$$

$$f_{(2)}^{X,11} = (-\tan\theta + \mu).\left[k_{12}.\tan^2\theta + 3k_{13}.\tan^3\theta.X_0 - 3k_{13}.\tan^2\theta.Y_0\right] + (1+\tan\theta)\left[k_{22} + 3k_{23}.X_0\right]$$
$$f_{(2)}^{X,12} = (-\tan\theta + \mu).\left[-2k_{12}.\tan\theta - 6k_{13}.\tan^2\theta.X_0 + 6k_{13}.\tan\theta.Y_0\right]$$
$$f_{(2)}^{X,22} = (-\tan\theta + \mu).\left[k_{12} - 3k_{13}.\tan\theta.X_0 - 3k_{13}..Y_0\right]$$

$$f_{(2)}^{Y,11} = -k_{12}.\tan^2\theta - 3k_{13}.\tan^2\theta.Y_0 + 3k_{13}.\tan^3\theta.X_0$$
$$f_{(2)}^{Y,12} = 2k_{12}.\tan\theta + 6k_{13}.\tan\theta.Y_0 - 6k_{13}.\tan^2\theta.X_0$$
$$f_{(2)}^{Y,22} = -k_{12} - 3k_{13}.Y_0 - 3k_{13}.\tan\theta.X_0$$

$$f_{(3)}^{X,111} = (-\tan\theta + \mu).k_{13}.\tan^3\theta + k_{23}.(1+\tan\theta) \qquad f_{(3)}^{Y,111} = k_{13}.\tan^3\theta$$
$$f_{(3)}^{X,112} = -3k_{13}.\tan^2\theta(-\tan\theta + \mu) \qquad f_{(3)}^{Y,112} = -3k_{13}.\tan^2\theta$$
$$f_{(3)}^{X,122} = 3k_{13}.\tan\theta(-\tan\theta + \mu) \qquad f_{(3)}^{Y,122} = 3k_{13}.\tan\theta$$
$$f_{(3)}^{X,222} = -k_{13}.(-\tan\theta + \mu) \qquad f_{(3)}^{Y,222} = -k_{13}$$

## APPENDIX C : FOURIER COEFFICIENTS OF NON-LINEAR FUNCTION AND THEIR DERIVATIVES

The $k$ -incremental vector of Fourier coefficients are arranged as follows:

$$\left\{Z^k\right\}=\left[\left\{Z^k_{1,0},\ldots,Z^k_{N,0}\right\}^T,\ldots,\left\{Z^k_{1,2j-1},\ldots,Z^k_{N,2j-1}\right\}^T,\left\{Z^k_{1,2j},\ldots,Z^k_{N,2j}\right\}^T,\ldots,\left\{Z^k_{1,2M},\ldots,Z^k_{N,2m}\right\}^T\right]^T$$

We note $I=\begin{bmatrix}1 & 0\\ 0 & 1\end{bmatrix}$ and $O=\begin{bmatrix}0 & 0\\ 0 & 0\end{bmatrix}$. The matrices $[A]$ and $[J]$ are given by

$$A=\begin{bmatrix} O & & & & \\ & \begin{bmatrix} O & \omega I\\ -\omega I & O\end{bmatrix} & & & \\ & & \ddots & & \\ & & & \begin{bmatrix} O & j\omega I\\ -j\omega I & O\end{bmatrix} & \\ & & & & \ddots \\ & & & & & \begin{bmatrix} O & M\omega I\\ M\omega I & O\end{bmatrix}\end{bmatrix}$$

$$[J]=([\Gamma]\otimes[I]).\begin{bmatrix}\ddots & & \\ & \begin{bmatrix}\partial f_x/\partial x & \partial f_x/\partial y\\ \partial f_y/\partial x & \partial f_y/\partial y\end{bmatrix} & \\ & & \ddots\end{bmatrix}.\left([\Gamma]^{-1}\otimes[I]\right)$$

with $f(x,y)=\sum_{i=0}^{L}\sum_{j=0}^{L}n_{ij}.x^i.y^j \Big/ \sum_{i=0}^{M}\sum_{j=0}^{M}d_{ij}.x^i.y^j$ , we have :

$$\partial f/\partial x=\left[\left(i.\sum_{i=0}^{L}\sum_{j=0}^{L}n_{ij}.x^{(i-1)}.y^j\right)\times\sum_{i=0}^{M}\sum_{j=0}^{M}d_{ij}.x^i.y^j-\sum_{i=0}^{L}\sum_{j=0}^{L}n_{ij}.x^i.y^j\times\left(i.\sum_{i=0}^{M}\sum_{j=0}^{M}d_{ij}.x^{(i-1)}.y^j\right)\right]\Big/\left[\sum_{i=0}^{M}\sum_{j=0}^{M}d_{ij}.x^i.y^j\right]^2$$

$$\partial f/\partial y=\left[\left(j.\sum_{i=0}^{L}\sum_{j=0}^{L}n_{ij}.x^i.y^{(j-1)}\right)\times\sum_{i=0}^{M}\sum_{j=0}^{M}d_{ij}.x^i.y^j-\sum_{i=0}^{L}\sum_{j=0}^{L}n_{ij}.x^i.y^j\times\left(j.\sum_{i=0}^{M}\sum_{j=0}^{M}d_{ij}.x^i.y^{(j-1)}\right)\right]\Big/\left[\sum_{i=0}^{M}\sum_{j=0}^{M}d_{ij}.x^i.y^j\right]^2$$

The DFT from time to frequency domain is given by

$$\left[\Gamma_{ij}\right]=\begin{cases}1/N & for\quad i=1\\ 2/N.cos[(j-1)i\pi/N] & for\quad i=2,4,\ldots,N-1\\ 2/N.sin[(j-1)(i-1)\pi/N] & for\quad i=1,3,\ldots,N\end{cases}\quad \text{for}\quad j=1,2,\ldots,N$$

and from frequency time domain:

$$\left[\Gamma_{ij}\right]^{-1}=\begin{cases}1 & for\quad j=1\\ cos[(i-1)j\pi/N] & for\quad j=2,4,\ldots,N-1\\ sin[(i-1)(j-1)\pi/N] & for\quad j=1,3,\ldots,N\end{cases}\quad \text{for}\quad i=1,2,\ldots,N.$$

Using the DFT method, it is possible to determine $Z$ and $F^{NL}$, the Fourier coefficients of $z\{x\quad y\}$ and $f$ , respectively :

$$Z=\Gamma.z$$
$$F^{NL}=\Gamma.f$$

| Harmonic coefficient | Case 1 $(M=1)$ | Case 2 $(M=2)$ | Case 3 $(M=3)$ |
|---|---|---|---|
| $Z_{1,0}$ | -0.685-0.0102i | -0.686-0.0102i | -0.685-0.0102i |
| $Z_{2,0}$ | -0.685+0.0102i | -0.686+0.0102i | -0.685+0.0102i |
| $Z_{1,1}$ | 1.7453+1.0786i | 1.7463+1.0812i | 1.7458+1.0808i |
| $Z_{2,1}$ | 1.7453-1.0786i | 1.7463-1.0812i | 1.7458-1.0808i |
| $Z_{1,2}$ | -1.0775+1.7377i | -1.0805+1.739i | -1.0801+1.7385i |
| $Z_{2,2}$ | -1.0775-1.7377i | -1.0805-1.739i | -1.0801-1.7385i |
| $Z_{1,3}$ | 0 | 0.0171+0.0215i | 0.0166+0.0212i |
| $Z_{2,3}$ | 0 | 0.0171-0.0215i | 0.0166-0.0212i |
| $Z_{1,4}$ | 0 | -0.0147+0.0388i | -0.0144+0.0382i |
| $Z_{2,4}$ | 0 | -0.0147-0.0388i | -0.0144-0.0382i |
| $Z_{1,5}$ | 0 | 0 | 0.0002+0.0007i |
| $Z_{2,5}$ | 0 | 0 | 0.0002-0.0007i |
| $Z_{1,6}$ | 0 | 0 | -0.0003+0.0008i |
| $Z_{2,6}$ | 0 | 0 | -0.0003-0.0008i |

Table 1 : Values of the harmonic coefficients

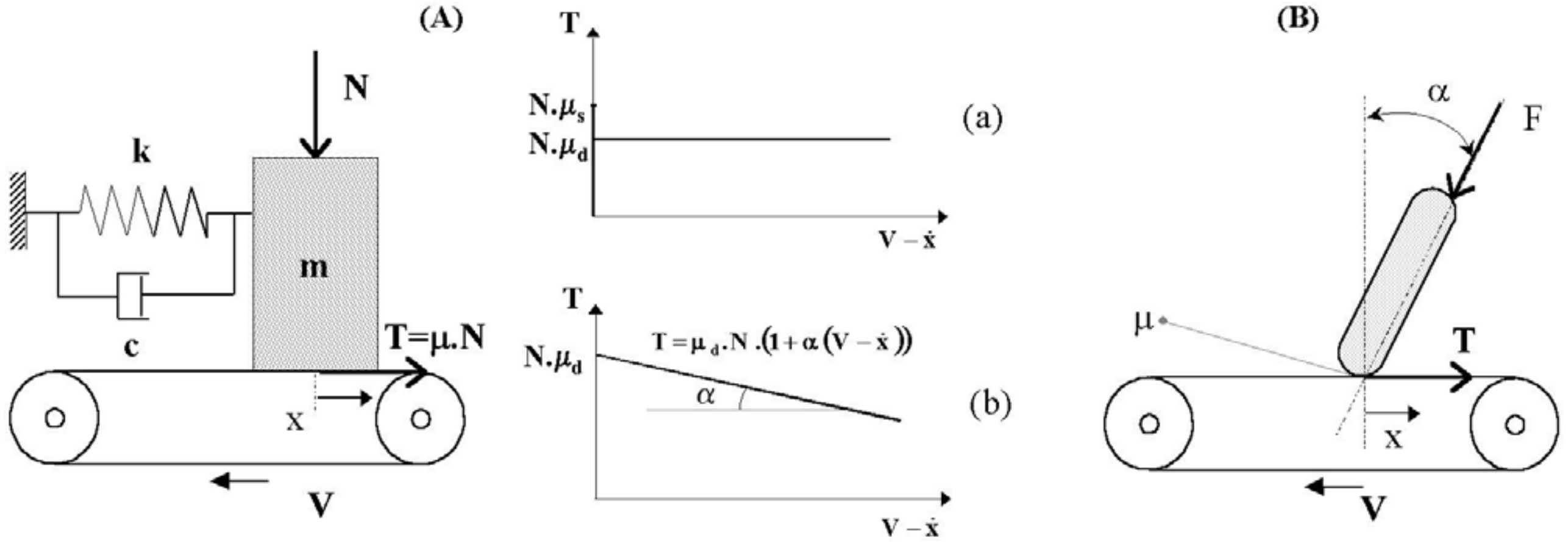

Figure 1 : Stick-slip and sprag-slip models

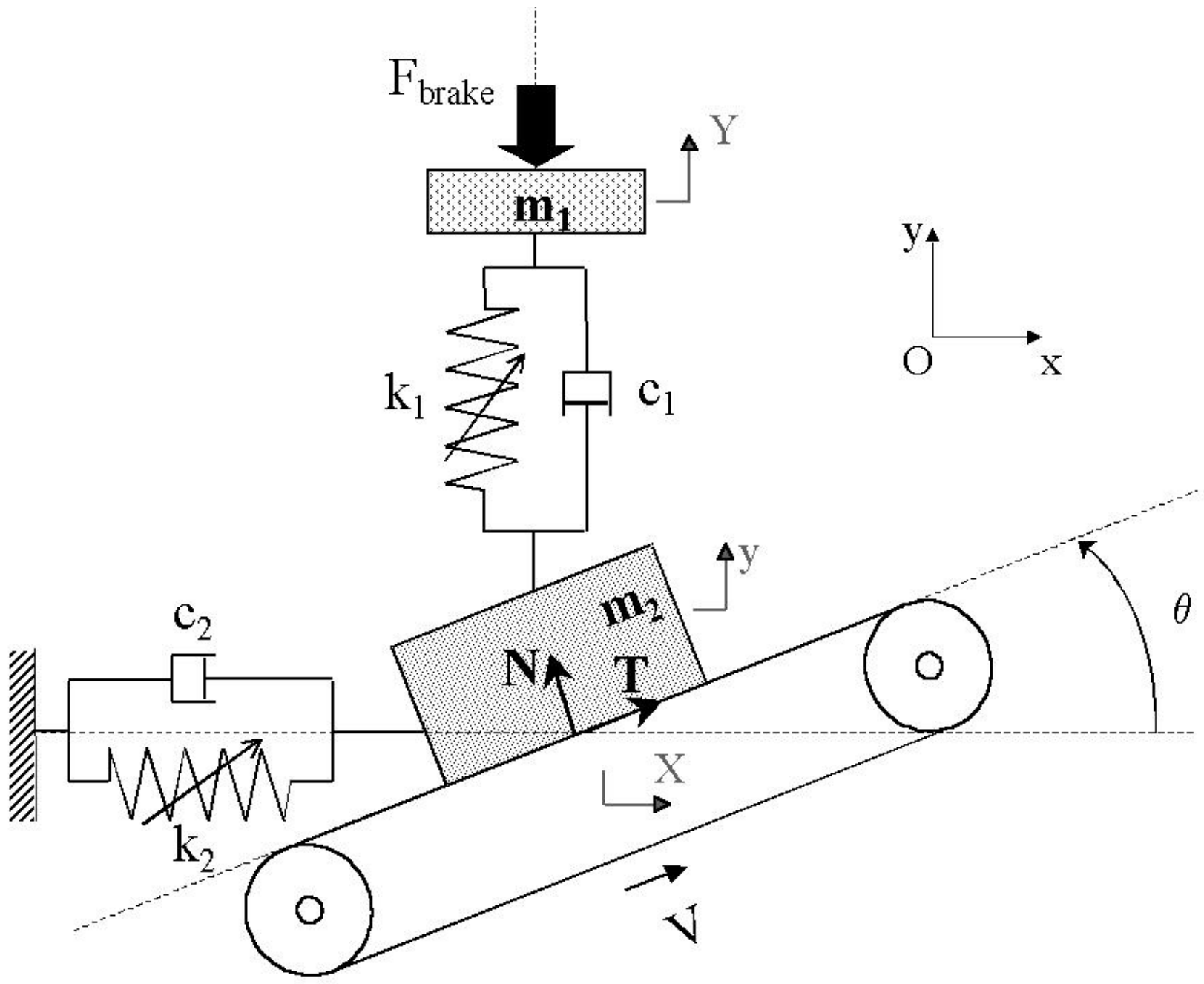

Figure 2 : Dynamic model of braking system

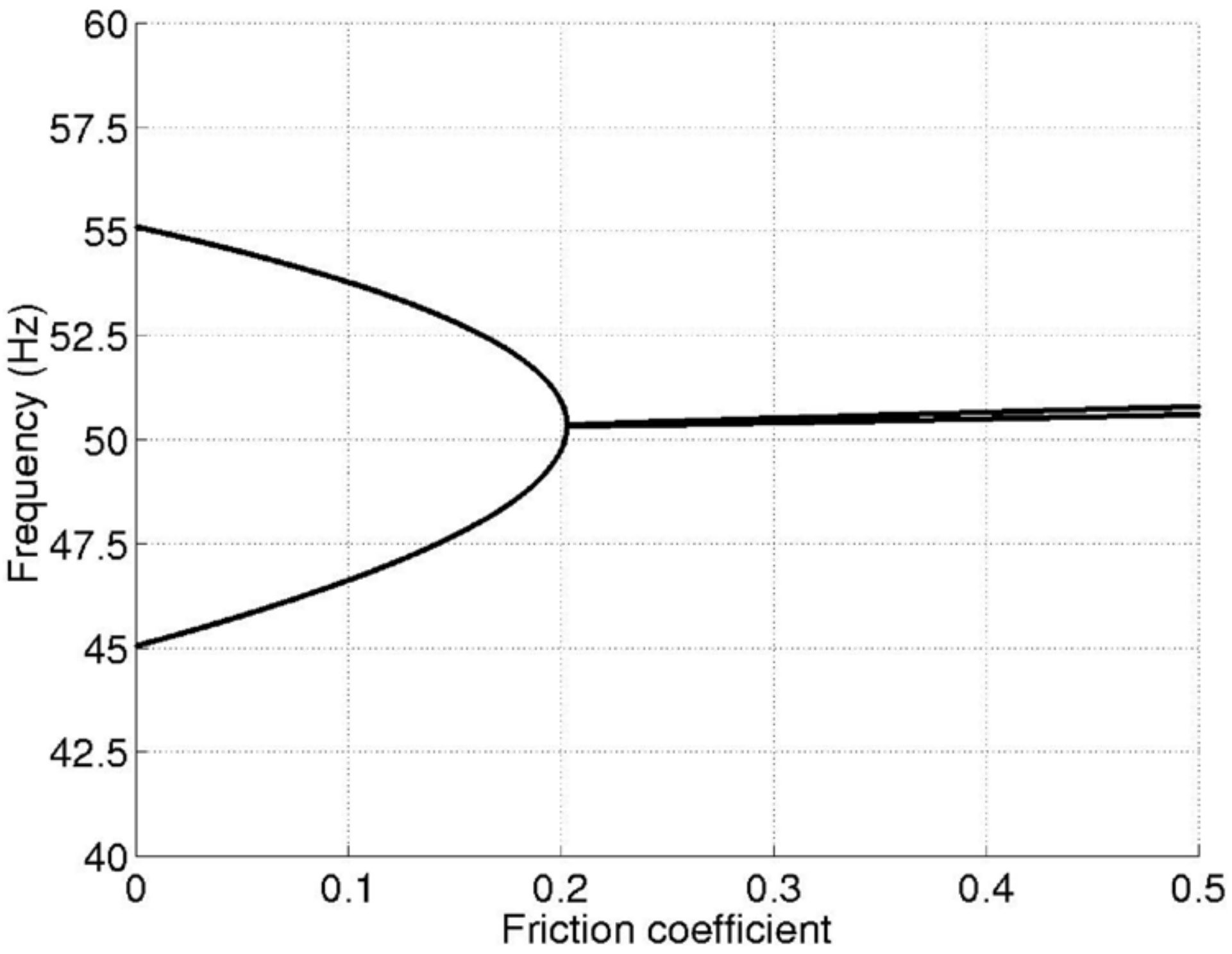

Figure 3 : Coupling of two eigenvalues

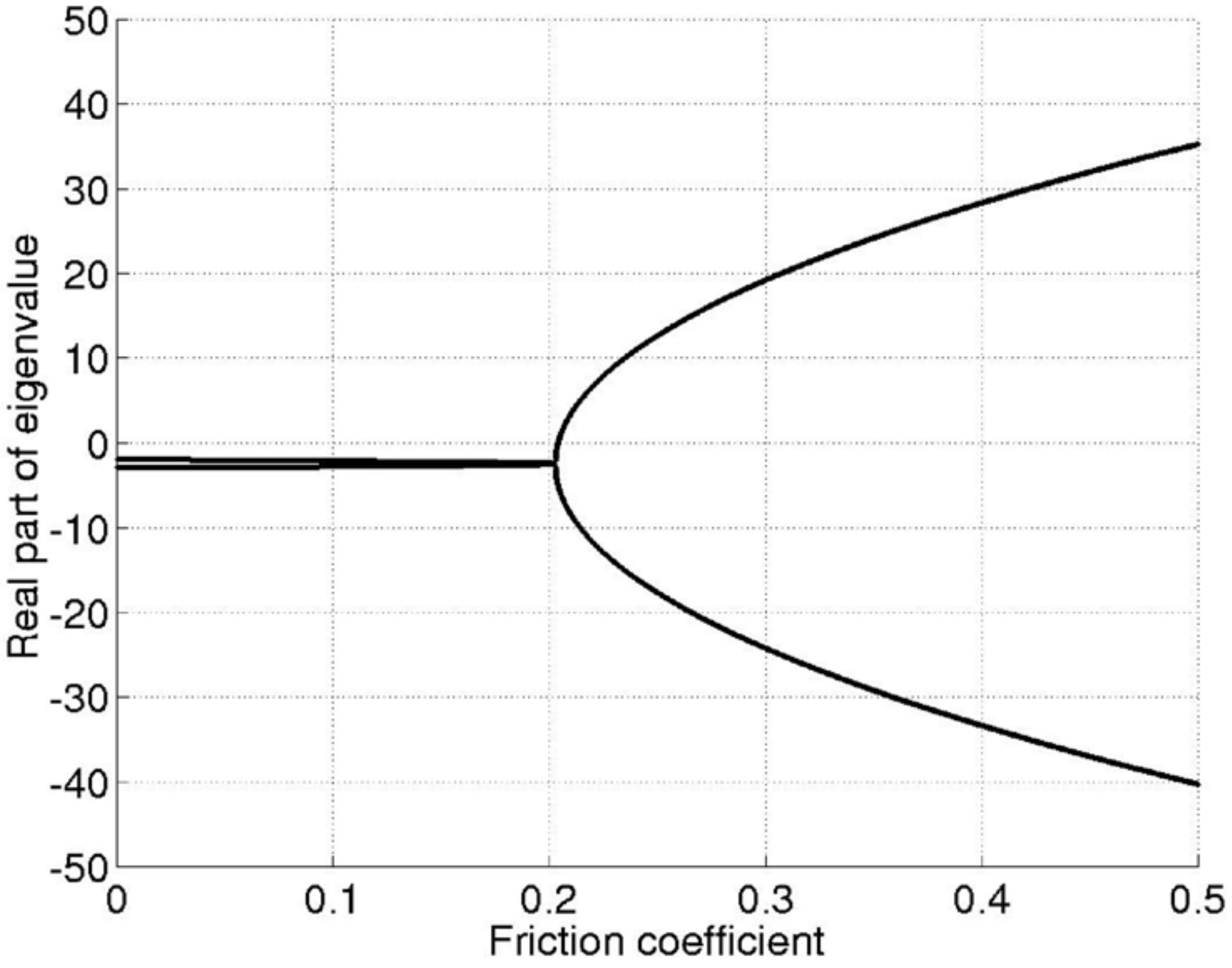

Figure 4 : Evolution of the real part of two coupling modes

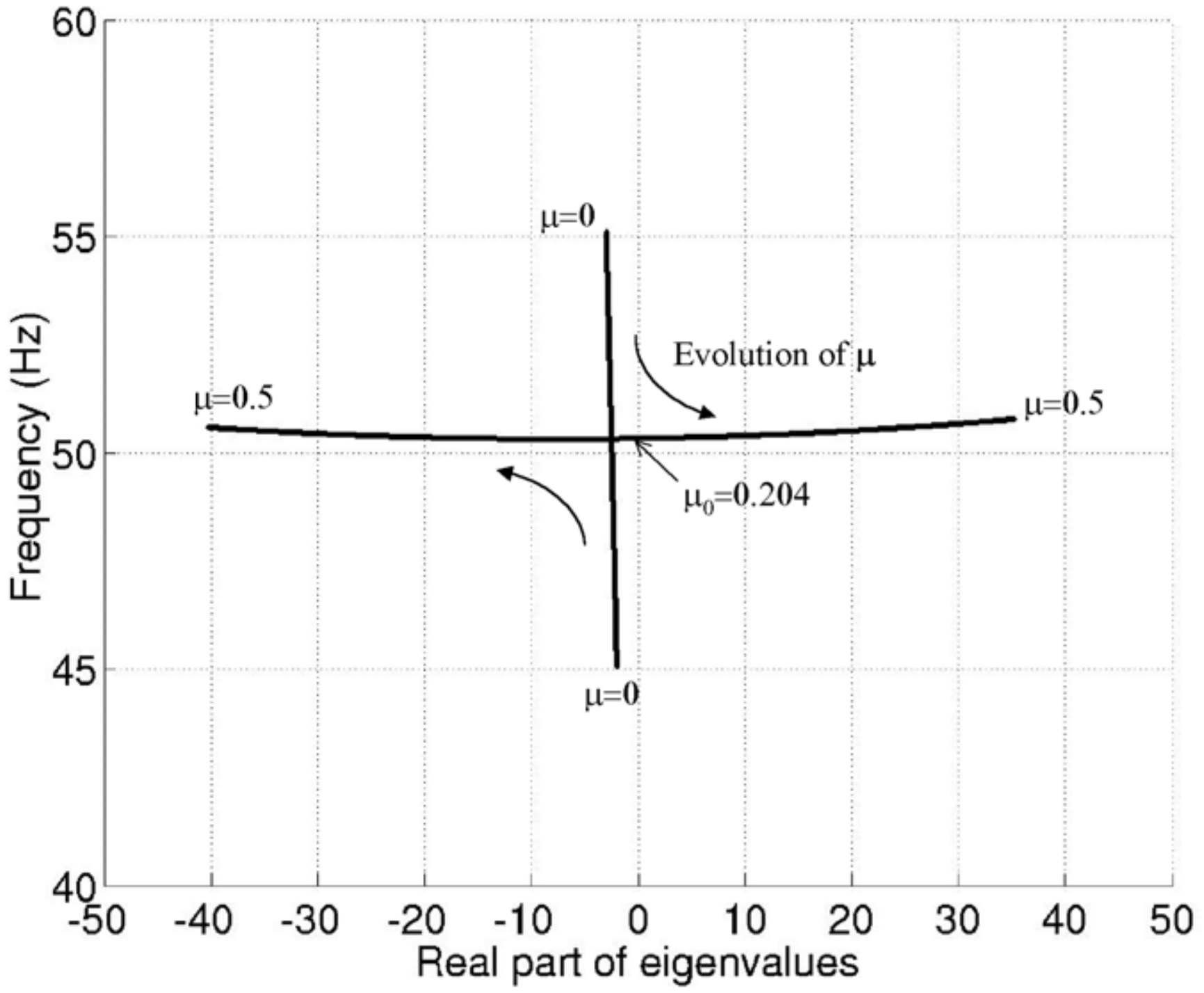

Figure 5 : Evolution of the eigenvalues by variation of $\mu$ in the complex plane

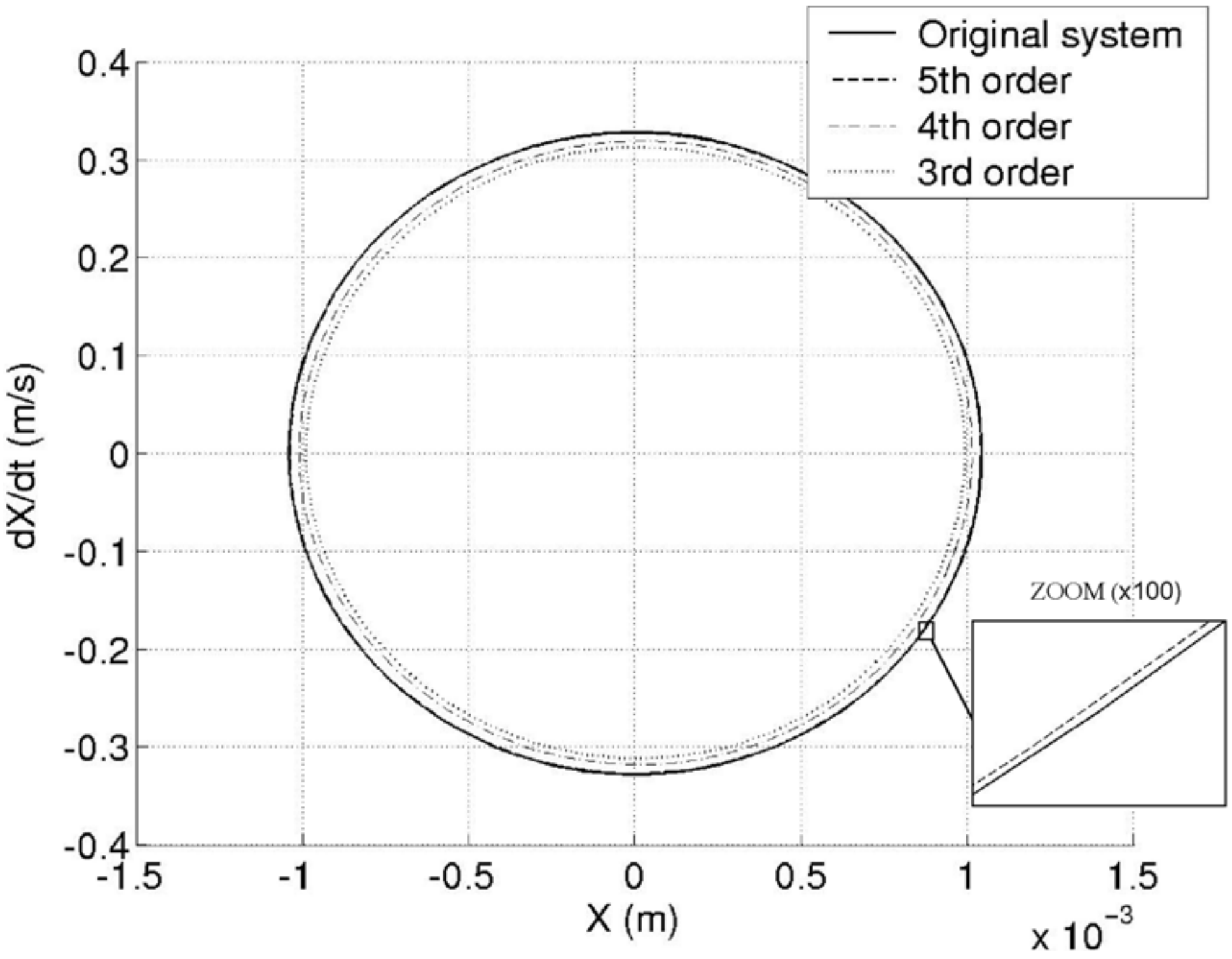

Figure 6 : X-limit cycle for $\overline{\mu} = 1/1000.\mu_0$

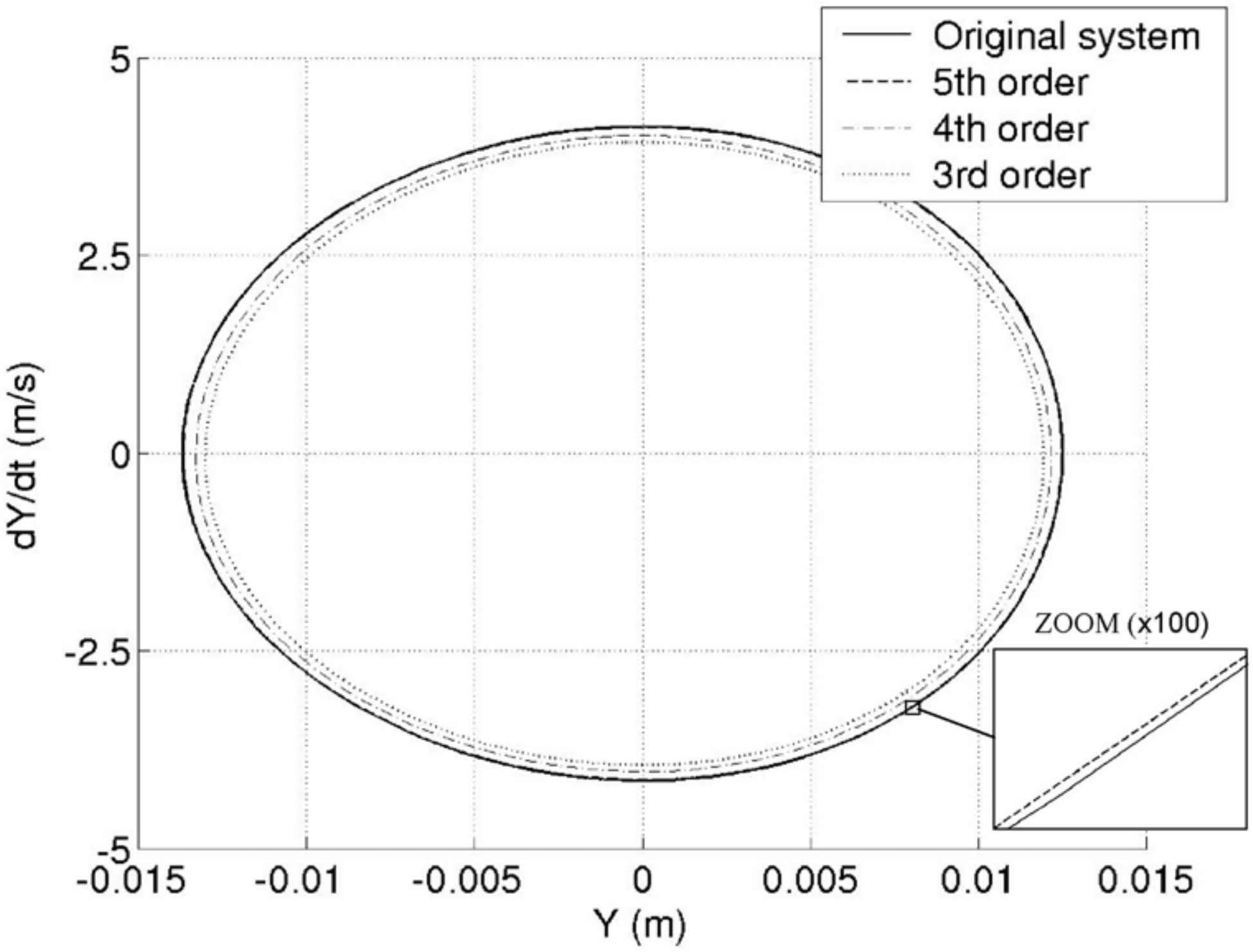

Figure 7 : Y-limit cycle for $\overline{\mu} = 1/1000.\mu_0$

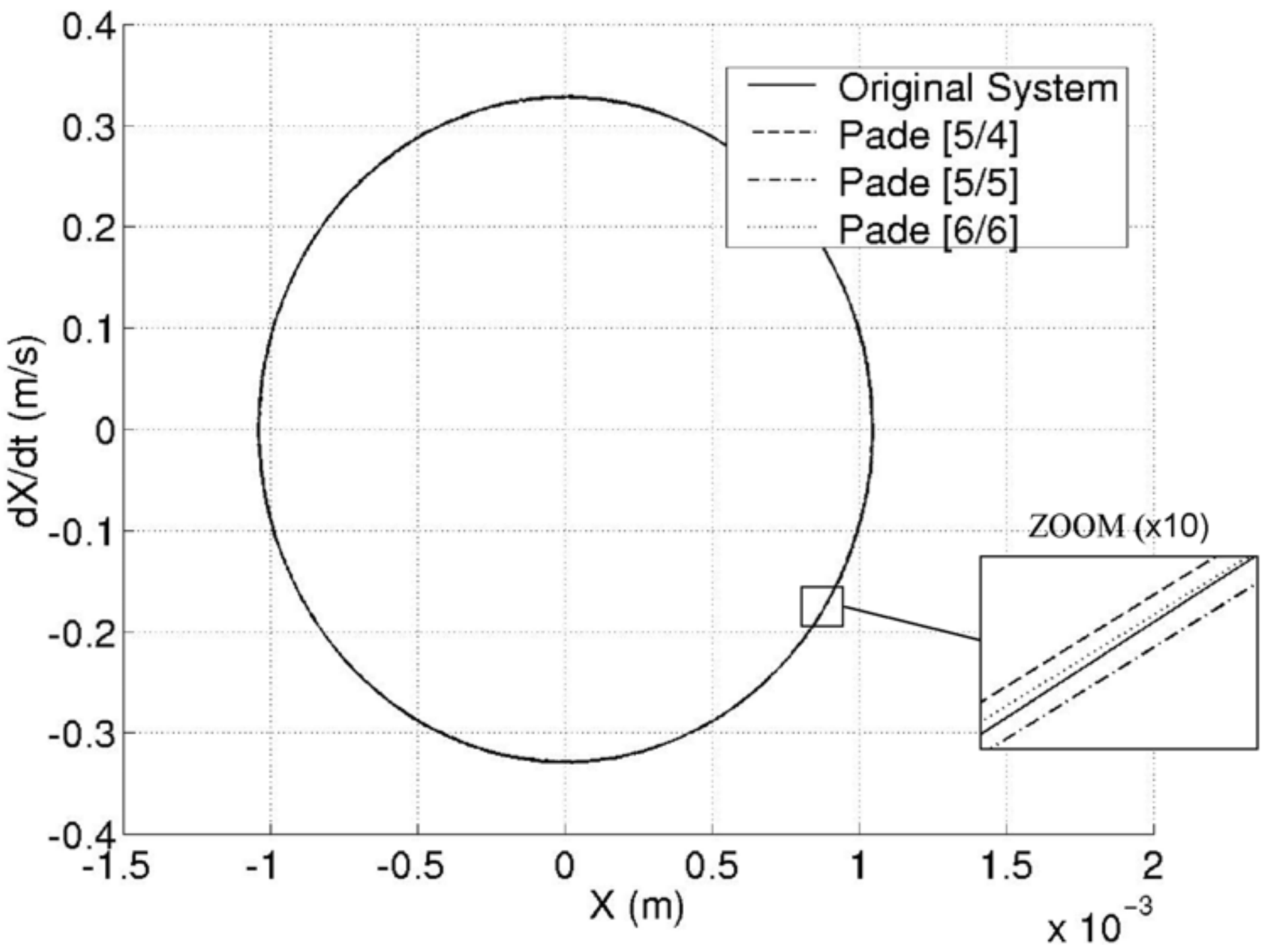

Figure 8 : X-limit cycle for $\bar{\mu} = \mu_0\ /\ 1000$

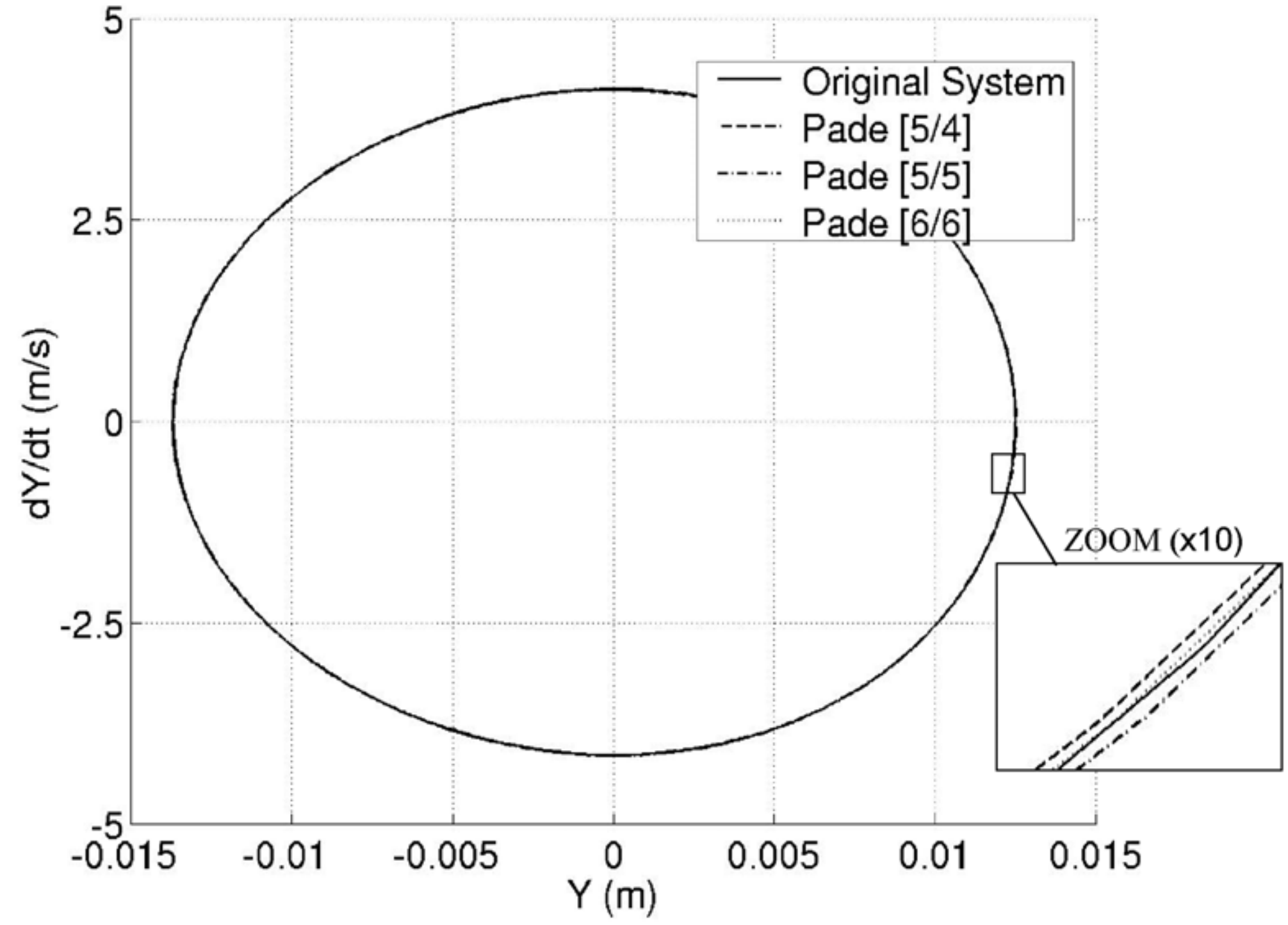

Figure 9 : Y-limit cycle for $\bar{\mu} = \mu_0\ /\ 1000$

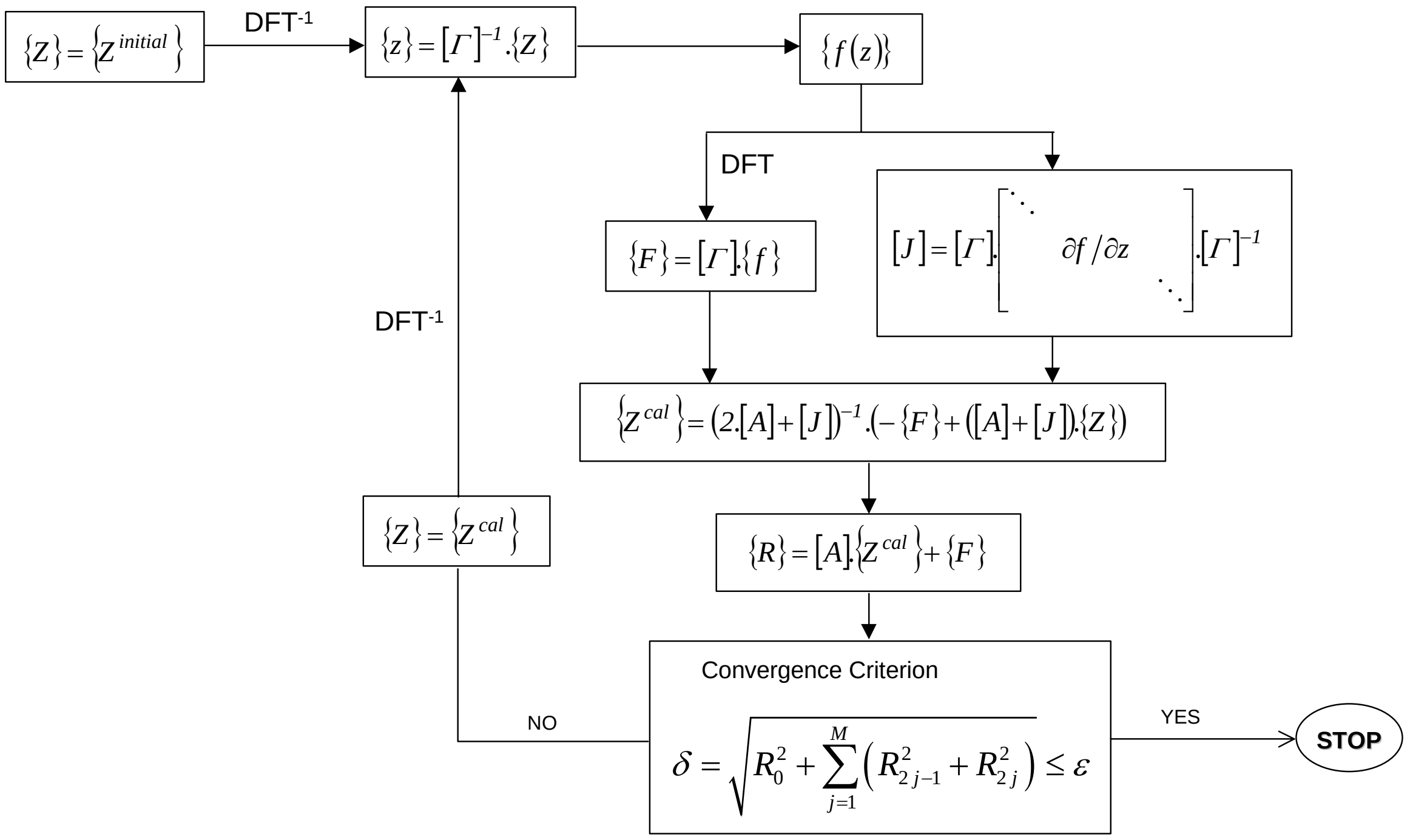

Figure 10 : Direct Iteration of the DFT method

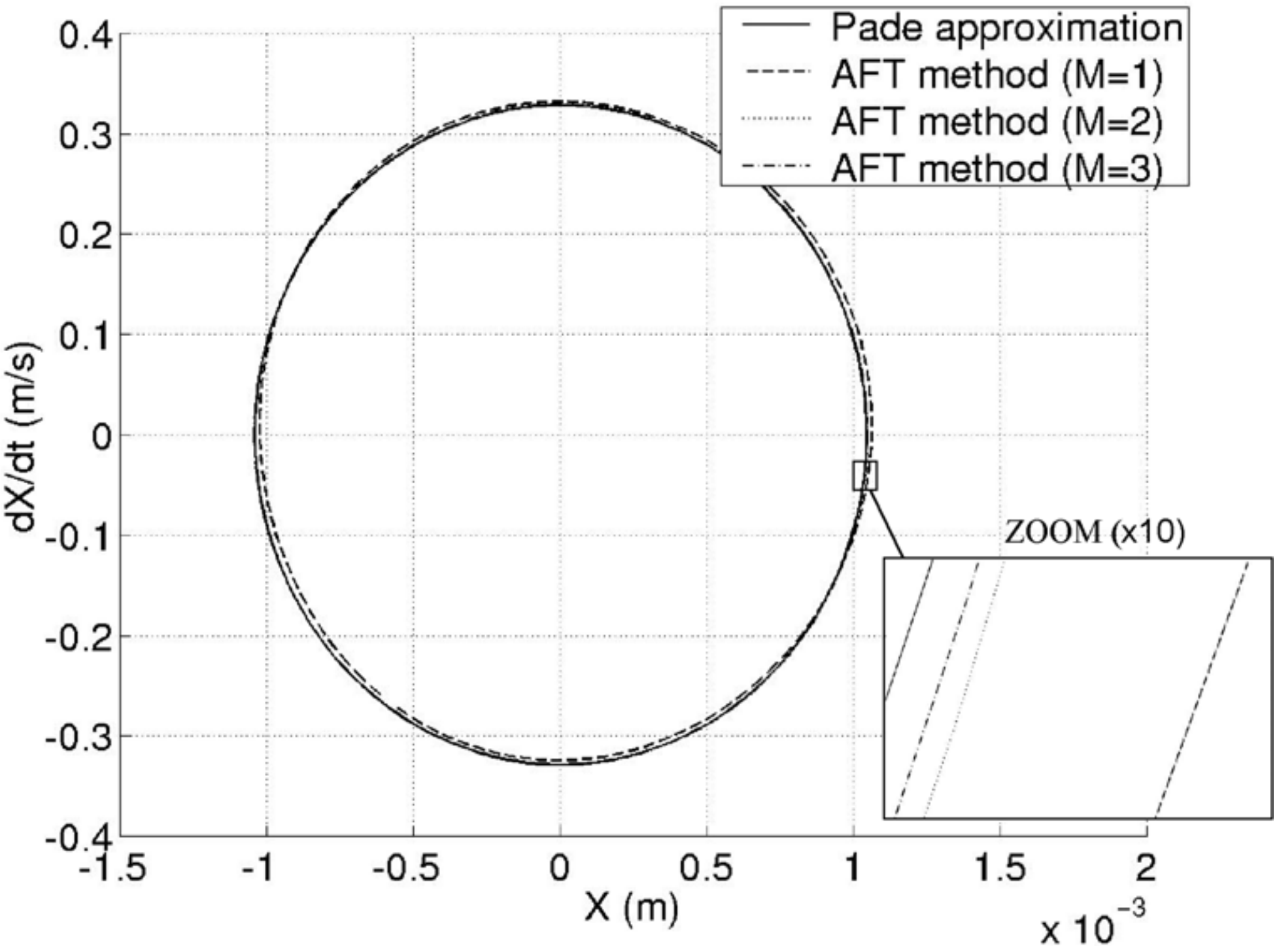

Figure 11 : X-limit cycles amplitude by using the alternate frequency/time domain method

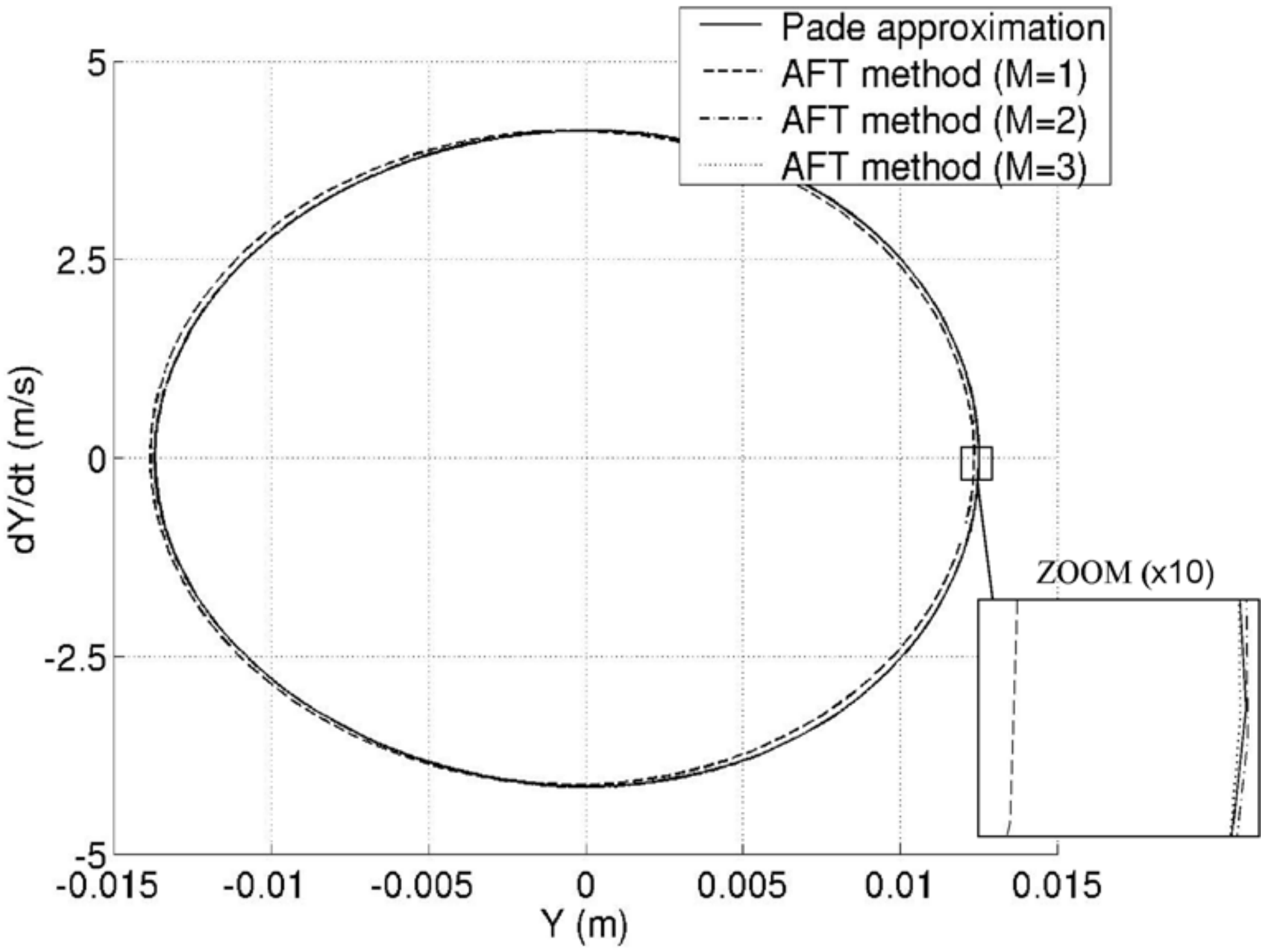

Figure 12 : Y-limit cycles amplitude by using the alternate frequency/time domain method